\documentclass[useAMS,usenatbib]{mnras}
\usepackage{setspace}
\usepackage{graphics}
\usepackage{graphicx}
\usepackage{amssymb}
\usepackage{longtable}
\usepackage{amsmath}
\usepackage{amsmath}
\usepackage{epstopdf}
\usepackage{color}
\usepackage[T1]{fontenc} 
\usepackage{aecompl}
\usepackage[none]{hyphenat}
\usepackage[normalem]{ulem}

\title[Formation of discs from infalling clouds onto SMBH binaries]{Infalling clouds onto super-massive black hole binaries --\\ I. Formation of discs, accretion and gas dynamics}
\author[Goicovic et al.]{F.G. Goicovic$^{1,3}$\thanks{E-mail:
fgarrido@astro.puc.cl}, J. Cuadra$^{1}$, A. Sesana$^{2,3}$, F. Stasyszyn$^{4}$, P. Amaro-Seoane$^{3}$ and \newauthor T.L. Tanaka$^{5,6,7}$ \\
$^{1}$ Instituto de Astrof\'isica, Pontificia Universidad Cat\'olica de Chile, Av.\ Vicu\~na Mackenna 4860, Santiago, Chile\\
$^{2}$ School of Physics and Astronomy, University of Birmingham, Edgbaston, Birmingham B15 2TT, United Kingdom\\
$^{3}$ Max-Plank-Institut f\"ur Gravitationsphysik, Albert Einstein Institut (AEI), Am M\"ulenberg 1, 14476, Potsdam-Golm, Germany\\
$^{4}$ Leibniz-Institut f\"ur Astrophysik Potsdam (AIP), An der Sternwarte 16, 14482, Potsdam, Germany	 \\
$^{5}$ Department of Physics and Astronomy, Stony Brook University, Stony Brook, NY 11794, USA \\
$^{6}$ Department of Physics, New York University, 4 Washington Place, New York, NY 10003, USA \\
$^{7}$ Max-Planck-Institut f\"ur Astrophysik, Karl-Schwarzschild-Str. 1, D-85741, Garching, Germany
}
\begin{document}

\date{\today}

\pagerange{\pageref{firstpage}--\pageref{lastpage}} \pubyear{2015}

\maketitle
%
%
%
\label{firstpage}

\begin{abstract}
There is compelling evidence that most --if not all-- galaxies harbour a super-massive black hole (SMBH) at their nucleus, hence binaries of these massive objects are an inevitable product of the hierarchical evolution of structures in the universe, and represent an important but thus-far elusive phase of galaxy evolution.
Gas accretion via a circumbinary disc is thought to be important for the dynamical evolution of SMBH binaries, as well as in producing luminous emission that can be used to infer their properties. One plausible source of the gaseous fuel is clumps of gas formed due to turbulence and gravitational instabilities in the interstellar medium, that later fall toward and interact with the binary.
In this context, we model numerically the evolution of turbulent clouds in near-radial infall onto equal-mass SMBH binaries, using a modified version of the SPH code \textsc{gadget}-3. 
We present a total of 12 simulations that explore different possible pericentre distances and relative inclinations, and show that the formation of circumbinary discs and discs around each SMBH (`mini-discs')  depend on those parameters.
We also study the dynamics of the formed discs, and the variability of the feeding rate onto the SMBHs in the different configurations.
\end{abstract}

\begin{keywords}
  accretion, accretion discs, black hole physics, hydrodynamics, galaxies: evolution, galaxies: nuclei
\end{keywords}

\section{Introduction}

The idea that super-massive black holes (SMBHs) reside in the nuclei of at least the more massive galaxies is currently well established \citep{rich98,Kormendy2013}. Additionally, according to our current structure formation paradigm, galaxies often interact and merge following the hierarchical growth of their parent dark matter halos. Then, in the aftermath of a major galaxy merger, we
expect that the merger product will contain a pair of SMBHs. Dynamical interaction with stars and gas drive the SMBHs towards the centre of the new galaxy on short time-scales, where they will form a gravitationally bound black hole binary \citep[BHB;][]{Beg80,MM01}.  The further evolution of the BHB will depend on the content and distribution of both gas and stars in the galactic nucleus \citep[see e.g.][]{Sesana2010,Khan2013,Holley2015}.

In gas-rich systems, the BHB evolution is likely to be driven primarily by gas. Many theoretical and numerical studies have focused on the orbital evolution of a sub-parsec binary surrounded by a gaseous circumbinary disc \citep{Ivanov1999,Gould2000,ArmNat05,C09,Haiman2009,Lod09,Nix11a,Roedig2011,Kocsis2012,Nix12,Roedig2012,Pau2013,Nix13,Roedig2014} 
and the electromagnetic signatures that such a system would produce \citep{Artymowicz1996,Milo2005,Tanaka2010,Tanaka2010b,	Tanaka2012,Sesana2012,DOrazio2013,Tanaka2013, Tanaka2013b, Roedig2014b,Brem2014, Farris2015}.
However, the exact mechanism that would form such discs is still unclear, as it critically depends on the fuelling processes onto those scales. 

Observational \citep[e.g.][]{Sanders1996} and numerical studies \citep[e.g.][]{May07} show an abundance of dense gas in the nuclei of merged galaxies. However, the wide range of physical scales involved makes the fuelling a very complex process -- the gas needs to lose its angular momentum efficiently to be transported from galactic scales down to the nuclear region \citep{Hopkins2011}. 
An important factor controlling the fuelling is the density distribution of the gas. Turbulence and gravitational instabilities in the interstellar medium can trigger the formation of gas clumps, which will be the seeds for molecular clouds \citep{Agertz2009}. Relaxation processes might then randomise the orbits of the clouds, producing discrete events of ``ballistic'' accretion onto a central SMBH, almost unaffected by hydrodynamical drag.

The large spread in angular momentum achieved with turbulence is likely to --at least partially-- randomise the direction of the accretion events even if the bulge has a net rotation \citep{Hobbs11}. A scenario in which SMBHs evolve by accreting individual gas clouds falling from uncorrelated directions has been proposed by \cite{KP06}. In later publications 
in the context of BHBs, Nixon and collaborators have shown that this scenario can lead to the formation of counter-rotating discs, facilitating the binary final coalescence \citep{Nix11a,Nix11b,Dunhill2014,Aly2015}. In a somewhat different application, the feeding of SMBHs through individual cloud infall events with different degree of angular momenta randomisation has been proposed as a viable scenario to reproduce current measurement of SMBH spins \citep{Dotti2013,Sesana2014}.

A perhaps more concrete example of this kind of accretion events is the putative cloud that resulted in the unusual distribution of stars orbiting our Galaxy's SMBH. Several numerical studies have shown that portions of a near-radial gas cloud infall can be captured by a SMBH to form one or more eccentric discs that eventually fragment to form stars \citep{BR08,HobNay09,Alig2011,Map12,Luc13},
roughly reproducing the observed stellar distribution \citep{Paumard2006, Lu2009}.

Working on the hypothesis that infalling clouds are common in post-merger galactic nuclei hosting BHBs, we study the formation and early evolution of discs in these events. Recently, \citet{Dunhill2014}  showed that misaligned clouds infalling  with a large impact parameter onto BHBs result in prograde or retrograde circumbinary discs.  In this study, we focus on cloud orbits with lower specific angular momentum, as well as higher inclinations relative to the BHB orbit, in order to investigate a wider variety of disc configurations and dynamics.

The paper is organised as follows. In Section \ref{sec_num}, we describe the details of our numerical model.
Section \ref{sec_discs} discusses the output of the simulations and the formation of discs depending on the inclination and impact parameter.
We detail in Section \ref{sec_accr} the accretion on to the binary and each SMBH, in particular its variability.
In Sections \ref{sec_mds} and \ref{sec_cbd}, we study the dynamics of the mini-discs and circumbinary discs, respectively. Finally, we discuss the implications of our results on the observability of these systems and future multi-messenger studies in Section \ref{sec_fin}, and summarise our study in Section~\ref{sec:summary}. This paper is the first on a series where we will be reporting the results of numerical models of infalling clouds onto BHBs; in upcoming publications we will study the orbital evolution of the binary, detailed observational signatures from the formed discs, and the impact of magnetic fields, among other topics.

\section{The numerical model}
\label{sec_num}

We model the interaction between the binary and cloud using a modified version of the smoothed particle hydrodynamics (SPH) code \textsc{gadget}-3, which is an improved version of the public code \textsc{gadget-2}  \citep{Springel2005}.

The binary is modelled by two equal-mass ``sink'' particles (see below) with initially circular, Keplerian orbits. The cloud is represented using approximately $4.2\times 10^6$ equal-mass SPH particles, with a total mass 100 times smaller than the binary. 

The code units are such that the initial semi-major axis,  mass and orbital period of the binary are unity. As a reference, this corresponds for one of our fiducial models presented in Section \ref{sec_fin} to a $10^6M_\odot$ binary, with a semi-major axis of $0.2$ pc and a period of $\approx 8400$ yrs.

\subsection{Initial conditions}

The cloud is initially spherical, with uniform density and a turbulent internal velocity field. The main effect of the turbulence is to provide support against the initial collapse due to the cloud's self-gravity. The velocities are drawn from a Gaussian random distribution with power spectrum $P_v(k)\equiv \langle |\vec v_k|^2\rangle \propto k^{-4}$, where $k$ is the wave number of the velocity perturbation, and the power index is chosen to match the observed velocity dispersion profile of molecular clouds \citep{Larson1981}.

To set the internal velocity field we treat the cloud as incompressible ($\vec\nabla\cdot \vec v=0$), which implies that we can represent the velocity field with a ``vector potential" $\vec A$.
We sampled the components of $\vec A$ in the Fourier space using an equi-spaced lattice with 256$^3$ coordinates. To assign a velocity vector to each particle, we inverse Fourier transform $\vec A$ and we interpolate the obtain values between grid points.
Finally, the velocity field is normalised such that the kinetic energy is equal to the absolute value of the potential energy, resulting in a cloud that is marginally unbound.

\begin{figure}
\centering
\includegraphics[width=0.46\textwidth]{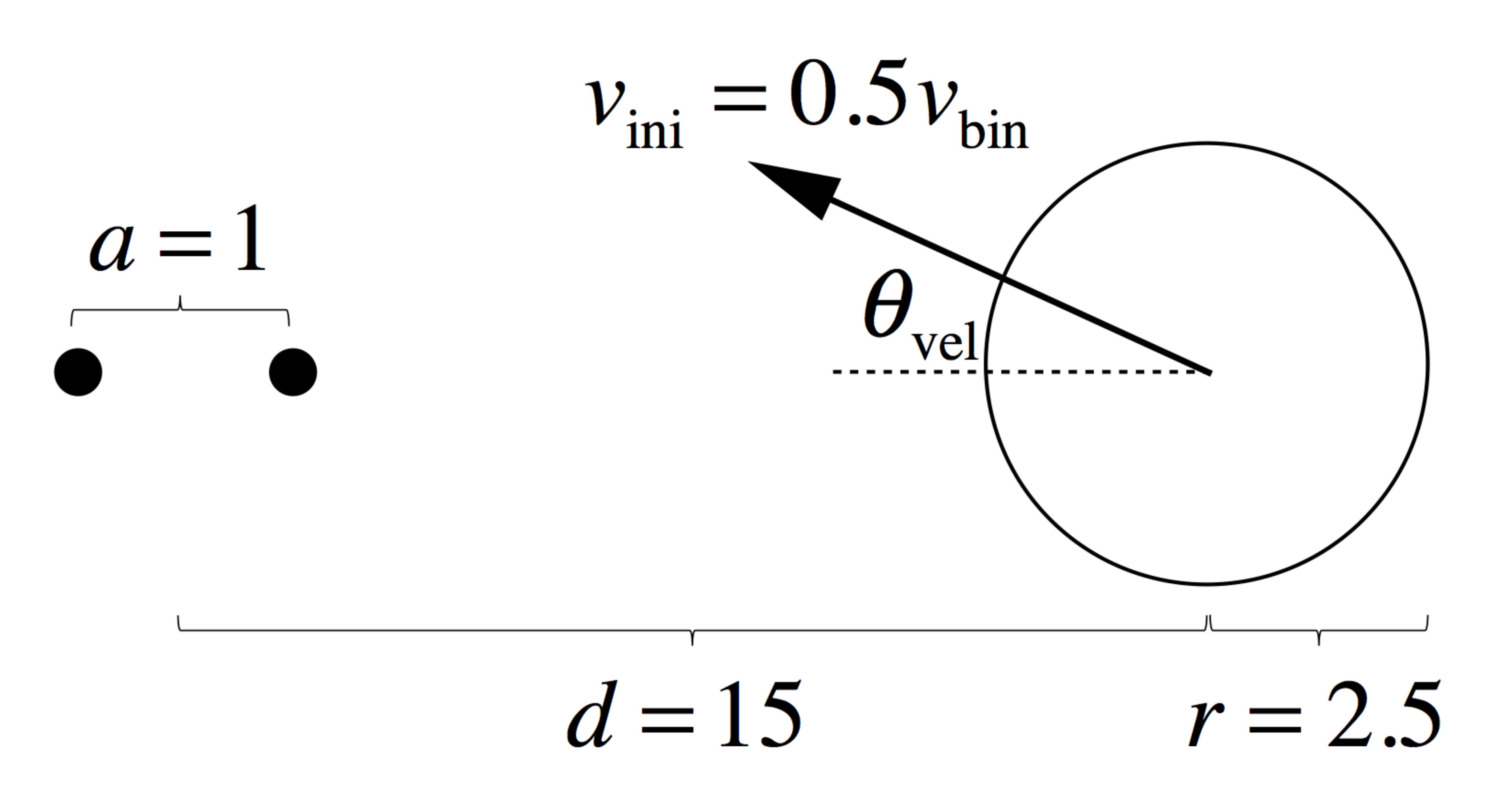}
\caption{Initial setup of the simulations. The circle on the right represents the cloud, while the small black, solid circles are the SMBHs. In our set of 12 simulations we sample different inclinations of the cloud initial velocity (i.e. different $\theta_{\rm vel}$) and orientations of the binary orbit.}
\label{initial}
\end{figure}

The physical setup of the simulations is shown in Figure~\ref{initial}. 
It is based on the work by \cite{BR08}, but replacing the central single black hole with a binary.
We place the cloud at a distance of 15 from the centre of mass of the binary, and we give it an initial velocity $\vec v_{\rm ini}$ such that it has an eccentric, bound orbit. The modulus of the initial velocity vector is constant in all our simulations $(v_{\rm ini}=0.5\,v_{\rm bin}$, where $v_{\rm bin}=0.5\sqrt{GM/a}$ is the tangential velocity of each SMBH), but we change its direction, reflected in the angle $\theta_{\rm vel}$. 
We model clouds with three different impact parameters; we choose $\theta_{\rm vel}=0.197$, $0.298$ and $0.403$ radians, so that the pericentre distances are $0.7$, $1.5$ and $3\,R_{\rm bin}$, respectively, where $R_{\rm bin}$ is the binary radius (i.e. half of the binary separation, $a/2=0.5$). 

Finally, since turbulence and relaxation might cause significant randomisation of the angular momenta of the gas clouds on parsec scales, we model clouds approaching with different relative orientations with respect to the binary orbit\footnote{Animations of our 12 simulations can be seen in \url{http://www.astro.puc.cl/~fgarrido/animations}}: 
\begin{itemize}
\item ``Aligned": the cloud starts in the plane of the BHB, with $\vec v_{\rm ini}$ lying on the same plane such that the cloud and the binary are co-rotating.
\item ``Counter-aligned": same as aligned, but counter-rotating.
\item ``Perpendicular edge-on": the cloud is initially in the same plane as the BHB, but the tangential component of $\vec v_{\rm ini}$ $(v_{\rm ini}\sin\theta_{\rm vel})$ is perpendicular to that plane.
\item ``Perpendicular face-on": the cloud starts in the plane perpendicular to that of the BHB, but the tangential component of $\vec v_{\rm ini}$ is parallel to that plane.
\end{itemize}

\subsection{Accretion}
\label{subs_accr}

The accretion recipe included in the standard version of  \textsc{gadget}-3 uses the Bondi model to estimate the amount of mass that should be added to each SMBH and then determines, probabilistically, the corresponding accreted particles \citep{Springel2005bh}. This prescription for the black hole growth is usually applied in cosmological simulations \citep[e.g.][]{Planelles2014} because the scales where the SMBHs are accreting are well below the resolution limit. In contrast, in our numerical models we do resolve the Bondi--Hoyle--Littleton radius, thus we shall use a more deterministic recipe for accretion.

In our simulations each black hole is represented by a 
``deterministic'' sink particle. That means, it accretes all SPH particles satisfying some given conditions within a certain radius \citep{Cuadra2006}. In addition, this type of particle interacts with others through gravity, but not with hydrodynamical forces.
In our model the SMBHs have a fixed accretion radius of $r_{\rm sink}=0.1$; each particle crossing $r_{\rm sink}$ is accreted if its kinetic energy is less than a fraction $\alpha$ of its potential energy\footnote{Both energies are computed in the reference frame of each SMBH.} \citep{Dotti2009}:
\begin{equation}
 E_{\rm kin}<\alpha |E_{\rm pot}|.
\end{equation}
Since our accretion radius is very large compared to the Schwarzschild radius\footnote{For a $10^6M_\odot$ binary separated 0.1 pc, the sink radius would be $\sim 10^5R_{\rm sch}$.} ($R_{\rm sch}$), this condition is necessary to avoid non-bound particles being added to the SMBHs. 
We adopt $\alpha=1$ throughout all the simulations presented here, meaning that all bound particles within the accretion radius are added to the corresponding SMBH. We did test $\alpha=0.5$ and also $r_{\rm sink}=0.05$, but the results were virtually unchanged, thus we kept the original values to save computational time. 
We discuss how the accretion rate measured onto the sink particles can be interpreted in Section \ref{sec_accr}.

The properties of the accreted particles, such as time, ID, mass, position and velocity, are stored once they are added to the SMBH \citep{Dotti2010}. Using this information we compute the total angular momentum of the gas accreted between outputs in the reference frame of the corresponding SMBH to explore the possibility of unresolved discs inside $r_{\rm sink}$ in Section~\ref{sec_mds}.

\subsection{Thermodynamics}

As we do not implement radiative cooling explicitly in our model, we use a barotropic equation of state, i.e. $P=P(\rho)$. The functional form of the pressure is chosen to mimic the thermodynamics -- in particular the temperature dependance with density -- of star-forming gas. 
At low densities the cloud is initially optically thin to the thermal emission from dust grains, and the compressional heating rate by the collapse is much smaller than the cooling rate by the thermal radiation. The situation reverses at high densities, when the compressional heating overwhelms the radiative cooling rate and the gas is heated as collapses \citep{Masunaga1998,Masunaga2000}.

The equation of state has the following form:
\begin{equation}
 P(\rho)=A(s)\rho^\gamma,
\end{equation}
where $P$ is the pressure, $\rho$ is the density, $A(s)$ is the entropic function \citep{Springel2002} and $\gamma$ is the polytropic index that depends on the density as follows:
\begin{equation}
\begin{split}
\gamma=1.0 \qquad\mbox{for}\qquad& \rho\leq \rho_c,\\
 \gamma=1.4 \qquad\mbox{for}\qquad& \rho > \rho_c,
\end{split}
\label{rhoc}
\end{equation}
with $\rho_c=1.096$ in code units. Note that the introduction of this two regime equation of state breaks the scale-free nature of our simulations. In Section~\ref{sec:scaling} we discuss the interpretation of scaling this number to different physical units. The index $\gamma=7/5$ corresponds to an adiabatic regime for diatomic gas and it is the value found by \citet{Masunaga2000} that represent the thermal evolution of a collapsing molecular cloud in the density regime we are modelling.

A consequence of using this equation of state is that we stop the collapse of the densest gas, avoiding excessively small time-steps that can stall the simulations. This type of equation of state is frequently used on hydrodynamical simulations of the evolution of turbulent clouds \citep[e.g.][]{Bate1995}. In the case of our model, this simple treatment of the thermodynamics allows us to capture the global behaviour of the gas during the interaction with the binary without an explicit implementation of cooling and/or radiative transfer. 
As the overall gas dynamics during the early phases of the interaction is dominated by the gravitational potential of the binary, the results presented here will depend only weakly on the thermodynamics adopted. However, the long-term evolution of the gas will be likely dependent on the thermodynamics, e.g. the cooling rate will determine whether the gas fragments or not. However the modelling of those processes is beyond the scope of this paper.

\section{Formation of discs}
\label{sec_discs}

\begin{figure}
\includegraphics[width=0.46\textwidth]{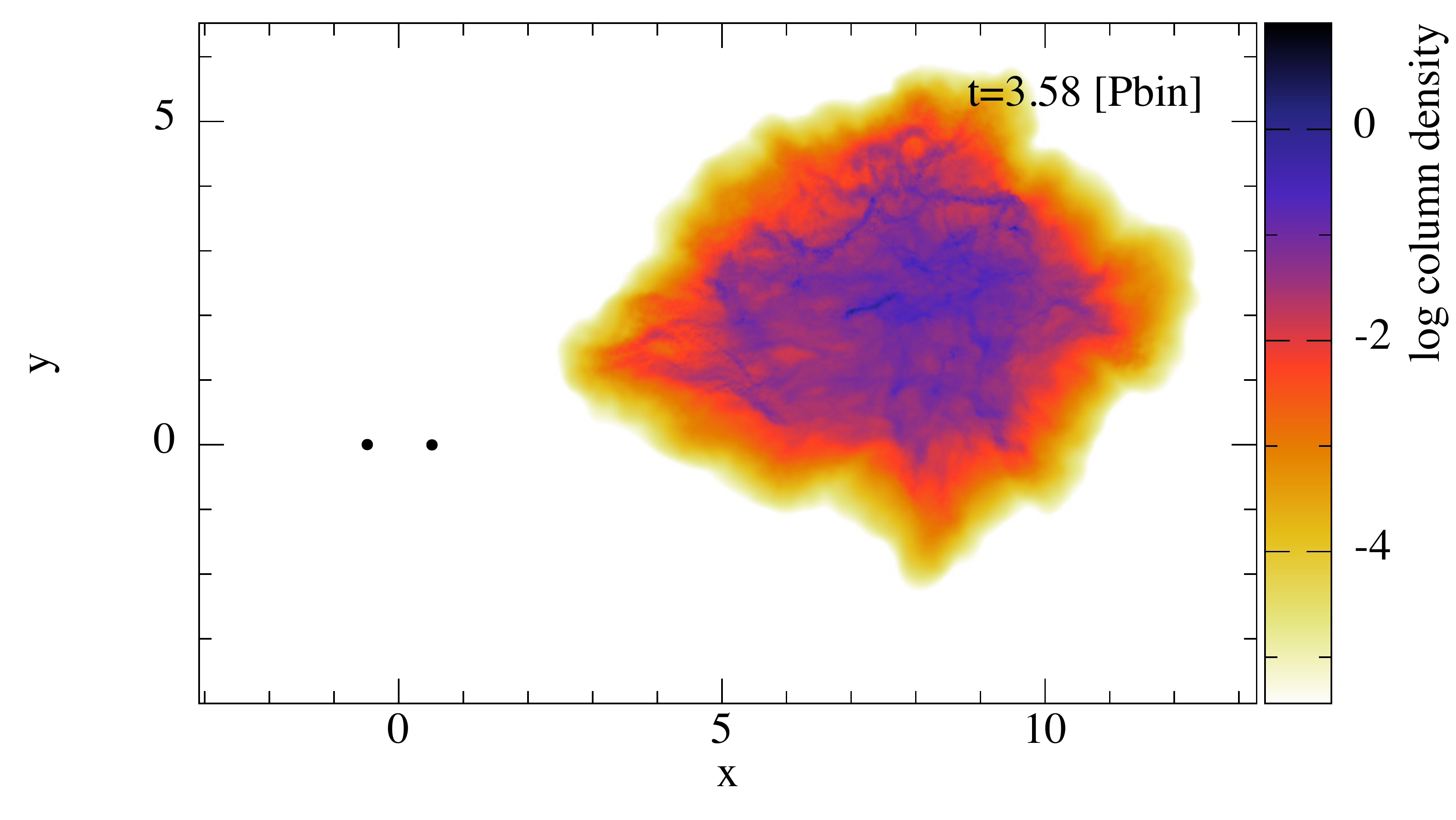}
\caption{Column density map of an early snapshot after the initial conditions of one of the simulations.
At this stage the cloud looks roughly the same for all orbital configurations, as its evolution is dominated by the initial turbulent velocity field.
The SMBHs are represented with two black circles on the left.}
\label{init}
\end{figure}

In Fig.~\ref{init} we show an early snapshot of one of our simulations as a column density map. Due to the large initial distance between the cloud and the BHB, the evolution of the gas is initially dominated by the turbulent velocity field, which produces filaments in the cloud. In this snapshot we  can already notice how the gas is stretched by the gravitational pull of the binary.  At this stage, the gas evolves almost independently of the particular orbital configuration of the system, because the effects that the BHB quadrupole potential can have on the hydrodynamics of the cloud are negligible, and the differences between impact parameters are small.

The study of the secular evolution of our systems, after the gas dynamics reaches a quasi-steady state, requires a considerable computing time that is not affordable with our standard configuration. We hence stop the simulations either once the transient effects of the cloud infall have stopped, or when the simulation stalls due to clump formation. We do however explore long-term effects with lower-resolution simulations in Section \ref{sec_cbd}. 

\begin{figure*}
\centering
 \begin{picture}(500,560)
  \put(10,0){\includegraphics[width=0.95\textwidth]{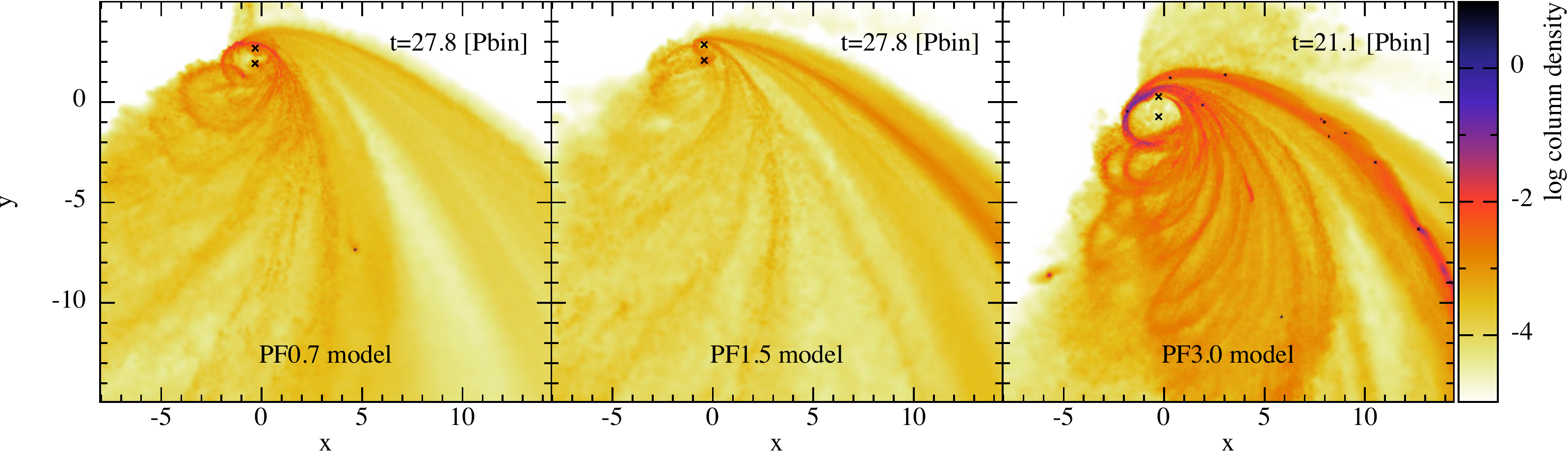}}
  \put(10,140){\includegraphics[width=0.95\textwidth]{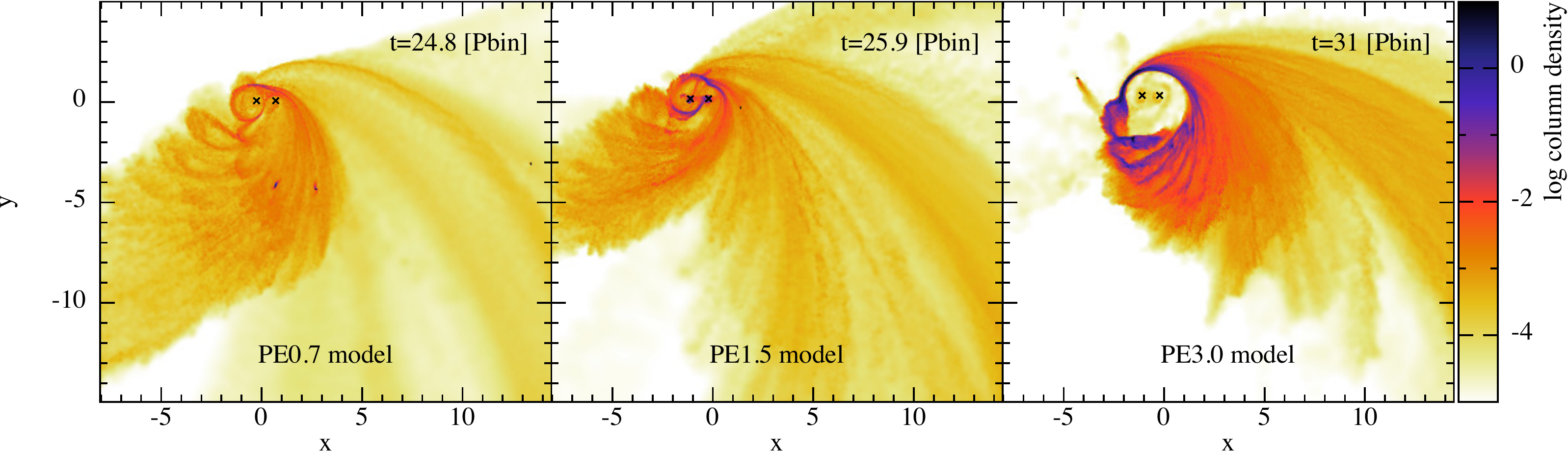}}
  \put(10,280){\includegraphics[width=0.95\textwidth]{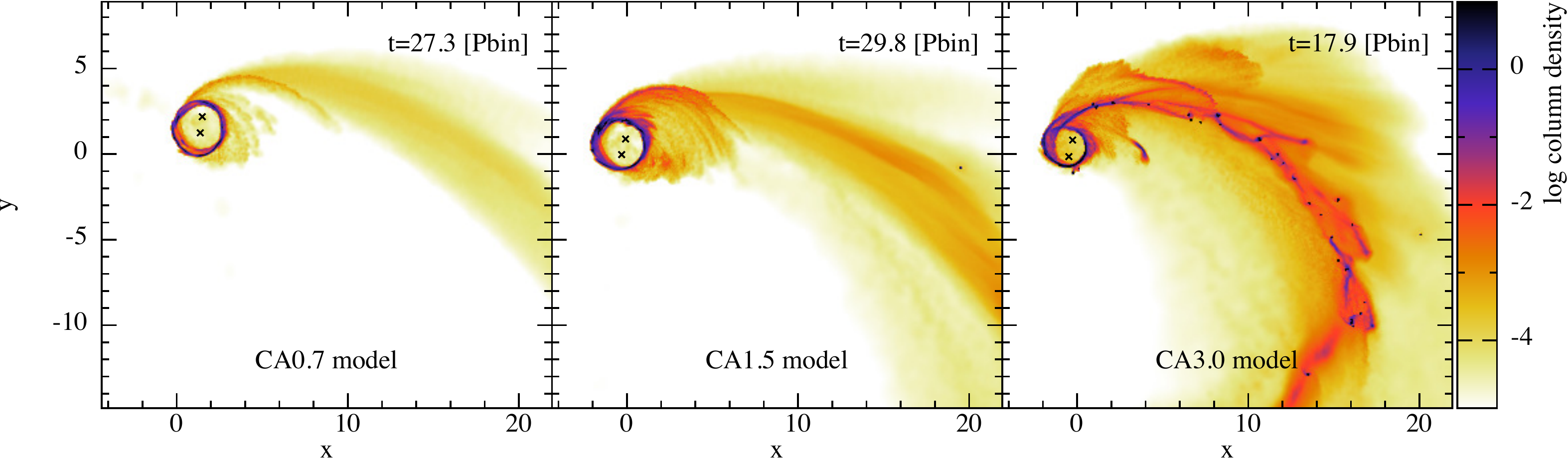}}
  \put(10,422){\includegraphics[width=0.95\textwidth]{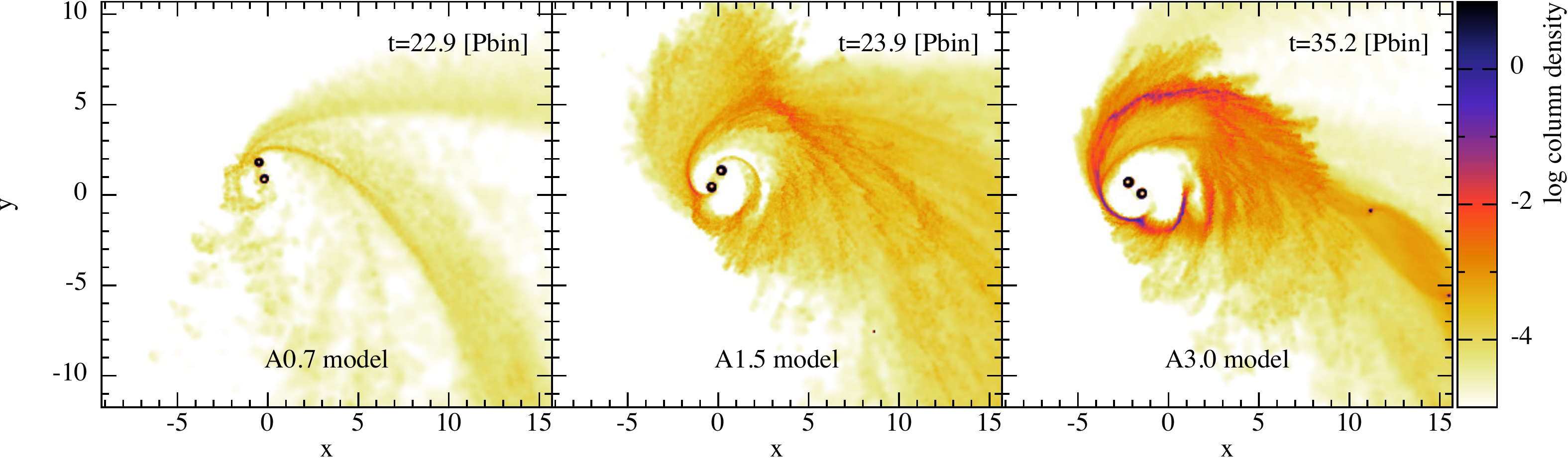}}
 \end{picture}
\caption{Column density maps of our simulations. The different rows are the 4 inclinations modelled, while increasing the impact parameter from left to right, as indicated by the model name. The position of the SMBHs is indicated by the black crosses. {\em Top row -- aligned (A) simulations:} The binary moves on the x-y plane, counter-clockwise. For every impact parameter dense ``mini-discs" form around each SMBH. As the impact parameter increases, we observe the formation of circumbinary discs. {\em Upper middle row -- counter-aligned (CA) simulations:} The binary moves on the x-y plane, clockwise. There is no formation of mini-discs. For each impact parameter we observe the formation of an almost circular and retrograde circumbinary ring. On the right panel we see that the gas fragments due to the compression during the first passage and forms a large amount of clumps. {\em Lower middle row -- perpendicular edge-on (PE) simulations:} The binary moves on the x-z plane. There is no formation of mini-discs. For the larger impact parameter the gas settles in bound orbits to form a circumbinary disc, which is clearly seen on the right panel. This disc is almost completely perpendicular to the orbit of the binary. {\em Bottom row -- perpendicular face-on (PF) simulations:} The binary moves on the y-z plane. Similar to the PE models, for the larger impact parameter the gas is able to form a perpendicular circumbinary disc.}
\label{BHB}
\end{figure*}

We show the final state of our aligned simulations (models A) in the top row of Fig.~\ref{BHB}. The first interaction of the gas with the binary is characterised by an efficient slingshot which pushes the remaining gas away from the system, shaping a tail. For the smallest impact parameter (top row, left panel of Fig.~\ref{BHB}), the strong interaction only allows the formation of the so-called ``mini-discs'' around each SMBH, while  increasing the impact parameter we allow more material to avoid the slingshot and settle in bound orbits around the binary, forming a circumbinary disc. The common feature of every impact parameter is the formation of mini-discs, appearing in the figure as black rings around either SMBH because of their high column density.  The discs tend to be slightly misaligned with respect to the binary orbit. The analysis of the dynamics of these mini-discs is shown in Section \ref{sec_mds}. {\em Only with this orientation we see the formation of extended, prominent and persistent mini-discs.} For the other BHB-cloud orientations, the material captured by individual SMBHs has little angular momentum, thus falling directly within $r_{\rm sink}$. Nevertheless, in the higher resolution tests shown in Section~\ref{subs_accr} we do observe the formation of {\em small and intermittent} mini-discs, which indicates that mini-discs might form on smaller scales, but cannot be resolved with our standard resolution. We explore the possibility of unresolved mini-discs inside the sink radius in Section~\ref{sec_mds}.

The final state of our counter-aligned simulations (models CA) is shown in the upper middle row of Fig.~\ref{BHB}. The interaction is very different because the typical gas velocity has the opposite direction relative to the orbital motion of the binary, cancelling most of the gas initial angular momentum. This enhances the accretion on to the BHB, as discussed in Section \ref{sec_accr}. The gas that remains bound after the interaction forms a very eccentric tail. For the three cases we observe the formation of a nearly circular, very narrow, counter-rotating circumbinary ring, arising as the material from the tail reaches the binary radius.  
The inner edge of these rings has a radius $\approx a$ from the center-of-mass (CoM) of the binary, which is expected due to the absence of resonances in a counter-rotating case \citep[e.g.][]{Nix11a}. The main qualitative difference observed on these three counter-aligned cases is the amount of gas clumps formed, as larger impact parameters result on more clumps.  Most of these clumps form due to the compression of gas during the pericentre passage that allows them to become self-gravitating, and eventually to form stars.  
We discuss the observational implications of star formation in Section \ref{sec_fin}.

We show the final state of our two perpendicular simulations in the lower rows of Figs.~\ref{BHB} (model PE, perpendicular edge-on and models PF, perpendicular face-on). As the typical gas velocity is perpendicular to that of the black holes, the slingshot is not very efficient, meaning that most of the gas is not pushed away from the binary. As we increase the impact parameter, we allow more material to settle on stable orbits. This gas forms a circumbinary disc, although completely misaligned with respect to the binary, keeping its initial angular momentum direction. In these configurations we also have regions of over-density which might lead to stellar formation, although they are located outside the region shown in the figures.

\begin{table*}
\centering
\caption{Properties of the discs formed in our simulations. In the model names the letter indicates the orbit orientation (A: aligned, CA: counter-aligned, PE: perpendicular edge-on, PF: perpendicular face-on) and the number the pericentre distance in units of binary radius. $t_{\rm fin}$ is the time at which we stop each simulation. $M_i$ and $r_{i,\rm{out}}$ $(i=1,2)$ are the mini-disc masses and outer radii, respectively. Available $M$ corresponds to the mass available to form a circumbinary disc, while $<e>$ is the median eccentricity of that gas.}
\label{table_discs}
\begin{tabular}{l*{8}{c}c}
\hline\hline
 Model & $t_{\rm fin}$ & Resolved & $M_1$ & $M_2$& $r_{\rm 1,out}$ & $r_{\rm 2,out}$ & Noticeable & Available $M$ & $<e>$\\
&$(P_{\rm bin})$ & mini-discs? &$(M_{\rm bin})$ &$(M_{\rm bin})$ & $(a)$& $(a)$&Circumbinary?& $(M_{\rm bin})$ &\\
\hline
 A0.7 & 22.9 & YES & $2.6\times10^{-4}$ & $1.4\times10^{-4}$ & 0.34 & 0.44 & NO & $2.9\times10^{-6}$ & 0.89 \\ 
 A1.5 & 23.9 & YES & $3.8\times10^{-4}$ & $4.3\times10^{-4}$ & 0.32 & 0.31 & YES & $1.1\times10^{-4}$ & 0.88  \\ 
 A3.0 & 35.2 & YES & $2.8\times10^{-4}$ & $4\times10^{-4}$ & 0.28 & 0.29 & YES & $1.3\times10^{-3}$ & 0.89  \\ \hline
 CA0.7 & 27.3 & NO & - & - & - & - & YES & $5.9\times10^{-4}$ & 0.16  \\ 
 CA1.5 & 29.8 & NO & - & - & - & - & YES & $2.6\times10^{-3}$ & 0.81  \\ 
 CA3.0 & 17.9 & NO & - & - & - & - & YES & $5.5\times10^{-3}$ & 0.83  \\ \hline
  PE0.7 & 24.8 & NO & - & - & - & - & NO & $1.5\times10^{-4}$ & 0.89 \\
 PE1.5 & 25.9 & NO & - & - & - & - & NO & $1.9\times10^{-3}$ & 0.91 \\ 
PE3.0 & 31.2 & NO & - & - & - & - & YES & $6.7\times10^{-3}$ & 0.75 \\ \hline
PF0.7 & 27.8 & NO & - & - & - & - & NO & $8.6\times10^{-5}$ & 0.91 \\ 
PF1.5 & 27.8 & NO & - & - & - & - & NO & $8.8\times10^{-4}$ & 0.97  \\ 
PF3.0 & 21.1 & NO & - & - & - & - & YES & $5.6\times10^{-3}$ & 0.91 \\  
\hline
\end{tabular}
\end{table*}

In Table \ref{table_discs} we compile different relevant values we measure for the discs formed in our simulations.  As noticed above, only the aligned cases show the formation of extended mini-discs, detectable at the resolution of our sink radius. These discs are very prominent and stable, and hence easy to identify. All their quantities presented have been computed using the median values over the last binary orbit. We observe that the disc masses are very similar for different impact parameters, which means that this quantity is -at least in this respect- independent of the amount of non-accreted gas. The inner edge of all these mini-discs extend from the accretion radius of our SMBH (see Section \ref{subs_accr}) to around the radius of the Hill sphere\footnote{$r_{\rm H}\approx0.3 a$ for a circular, equally-massive binary}, as expected.
 
The circumbinary discs are more difficult to analyse.  It is not possible to determine at the time we stop the simulations which fraction of the gas is going to form a stable disc.
Hence,  we first simply visually establish whether there is gas orbiting the binary in closed orbits -- this criterion is stated in Table~\ref{table_discs}. 
Additionally, to estimate the possible disc properties, we identify the gas particles that could become part of a circumbinary disc.  These are going to be the particles that are bound to the binary, and have orbits calculated around the CoM with pericentre distances $b$ larger than a threshold value defined to be $r_{\rm min}=2a$ for the aligned cases and $r_{\rm min}=a$ for the rest. Gas particles with a pericentre distance smaller than this radius will either be re-ejected in a slingshot process that prevents the gas to complete an orbit around the binary (see e.g. top row, left panel of Fig.~\ref{BHB}), or accreted, or will become part of one of the minidiscs, if present.  For all gas particles fulfilling the $b>r_{\rm min}$ criterion, we measure as before the total mass and median eccentricity from the last snapshot in each simulation.
The amount of gas available to form a circumbinary disc increases with the impact parameter, simply because gas with larger angular momentum avoids being accreted. The median eccentricity of the gas is high for most cases due to the initial orbit of the cloud: after the first passage most of the gas forms a very eccentric tail.
The exception is case CA0.7, here, due to the small pericentre distance, most of the gas is accreted, leading to a very light tail respect to the circumbinary ring. The other cases have a more massive tail, which make the eccentricity distribution skew towards high values.

In conclusion, the  interaction of individual clouds with a binary results in circumbinary discs that are initially very eccentric ($e\gtrsim0.8$, note that this estimate includes bound material within the highly eccentric tails, the long-term evolution of eccentricity will be discussed in Section \ref{sec_cbd}) . Moreover, by our choice of parameters, the disc masses are at most roughly half of the initial cloud mass, and in most cases much less than that, only a few percent. This does not only make them hard to detect directly, but also implies a small effect on the secular evolution of the binary.   This conclusion is likely to change, however, if we consider the cumulative effect of many cloud infall events, or the effect of individual massive clouds.  The results of that study will be presented in a follow-up paper.

\section{Accretion rate and total accreted mass}
\label{sec_accr}

\begin{figure*}
\centering
 \begin{picture}(500,520)
  \put(0,0){\includegraphics[width=0.3333\textwidth]{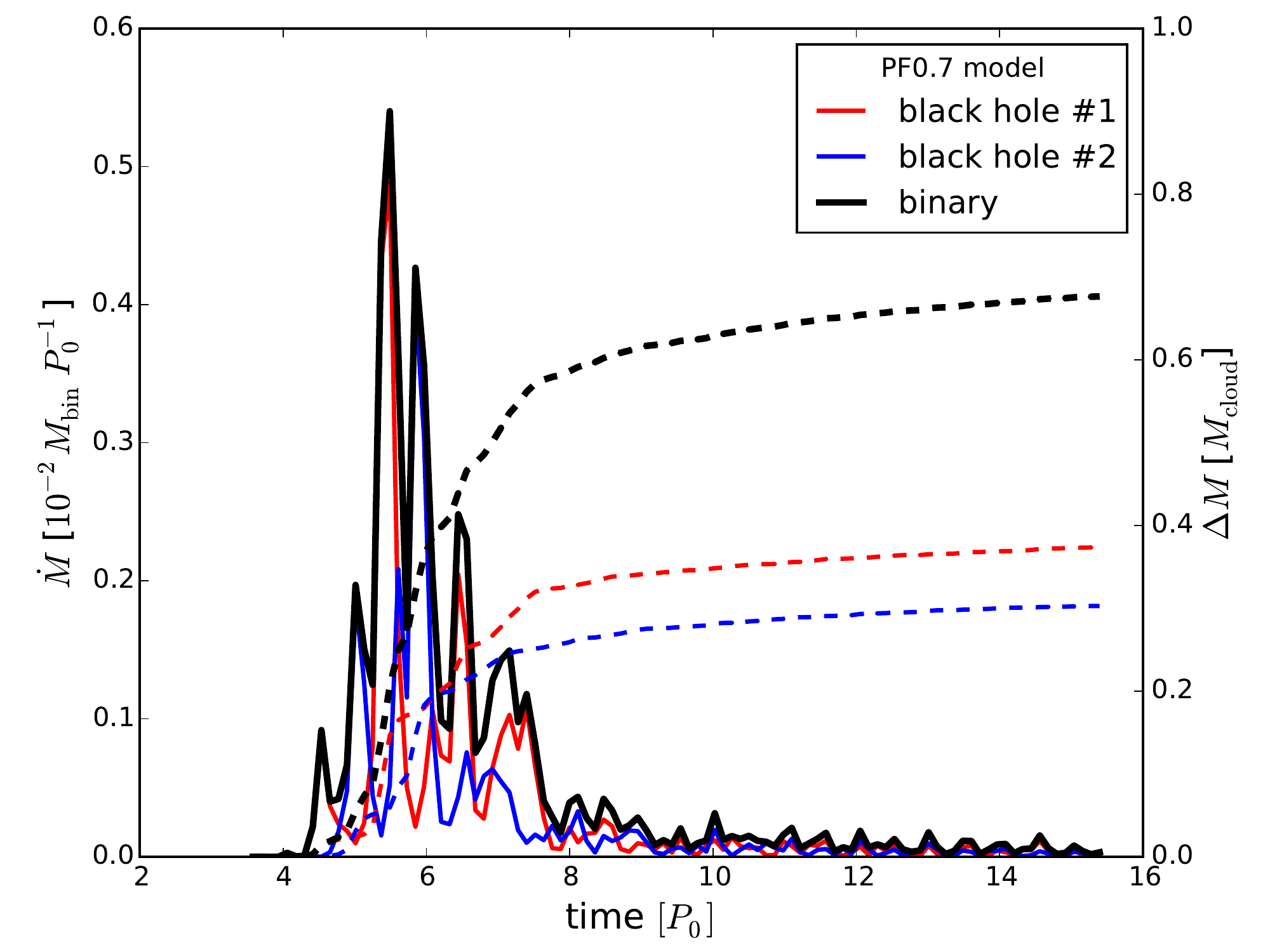}}
  \put(167,0){\includegraphics[width=0.3333\textwidth]{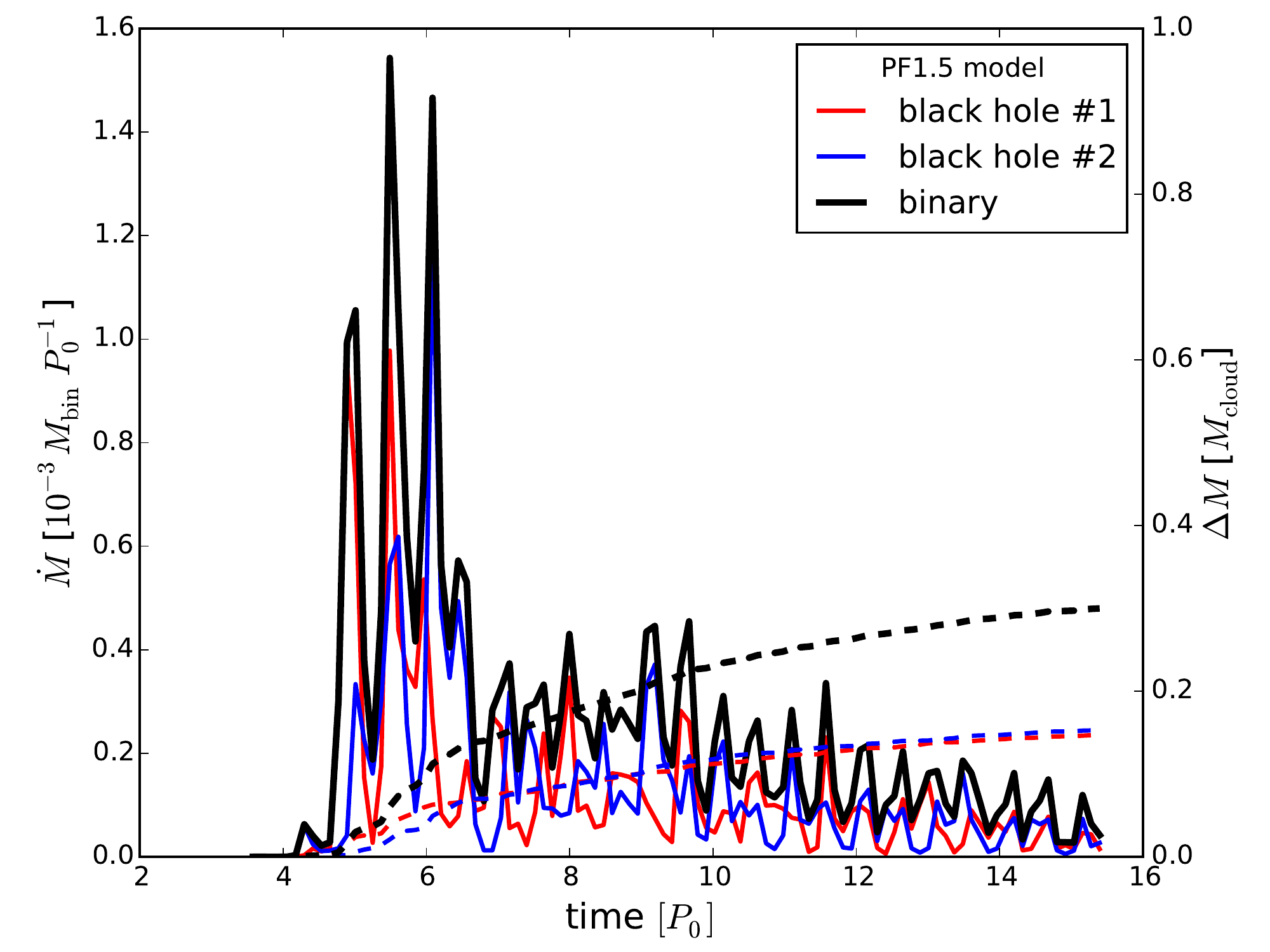}}
  \put(333,0){\includegraphics[width=0.3333\textwidth]{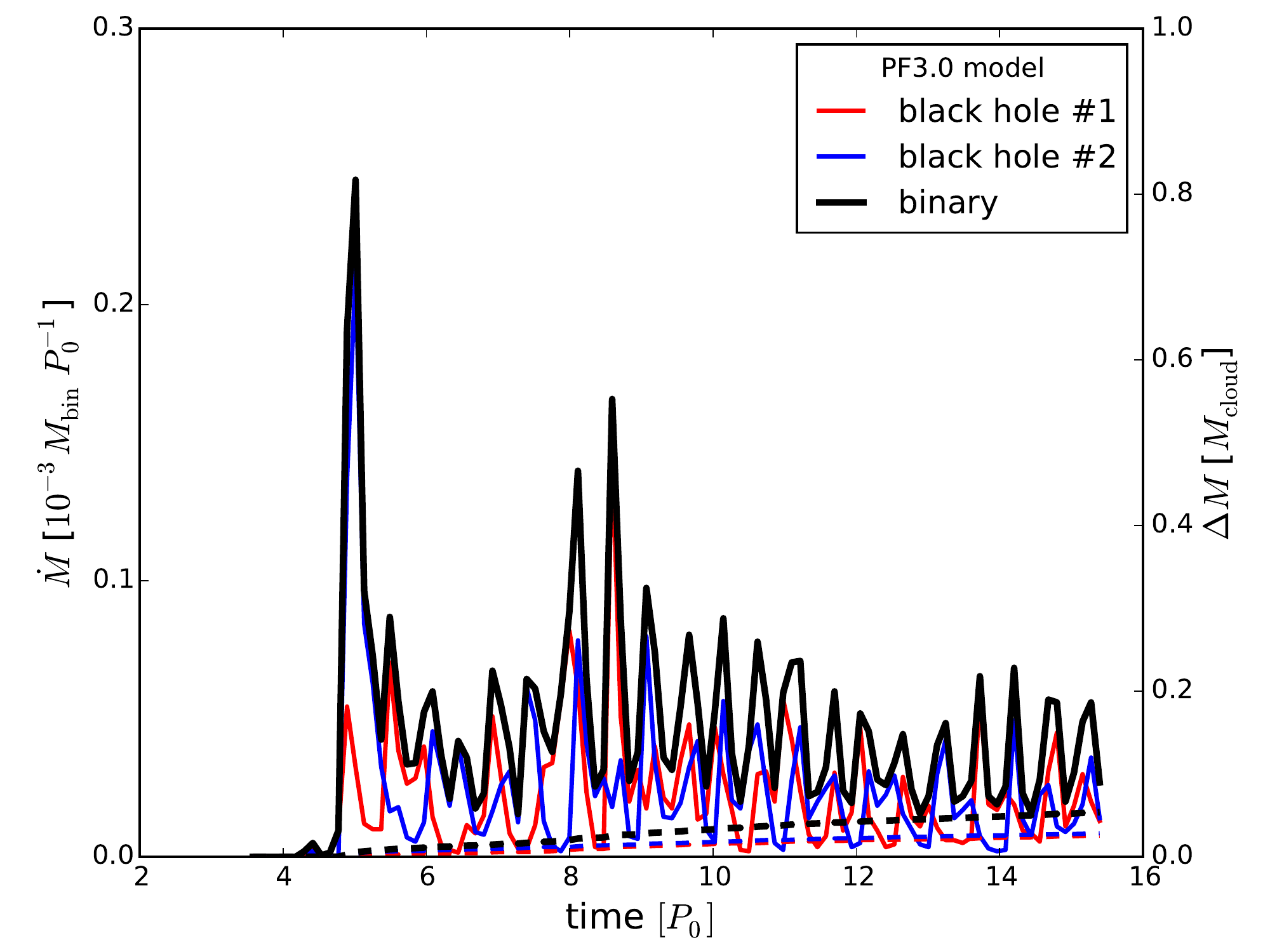}}
  \put(0,130){\includegraphics[width=0.3333\textwidth]{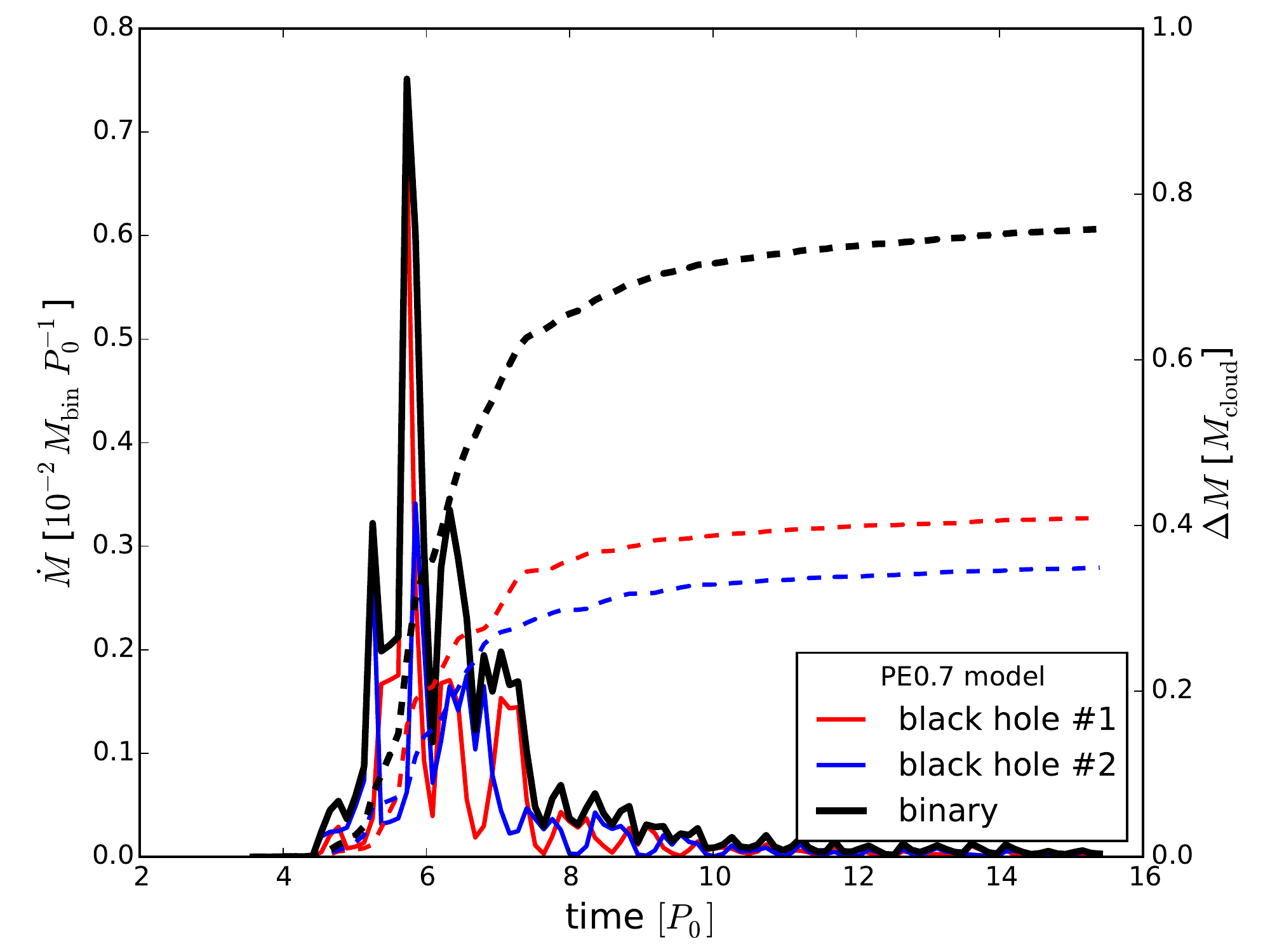}}
  \put(167,130){\includegraphics[width=0.3333\textwidth]{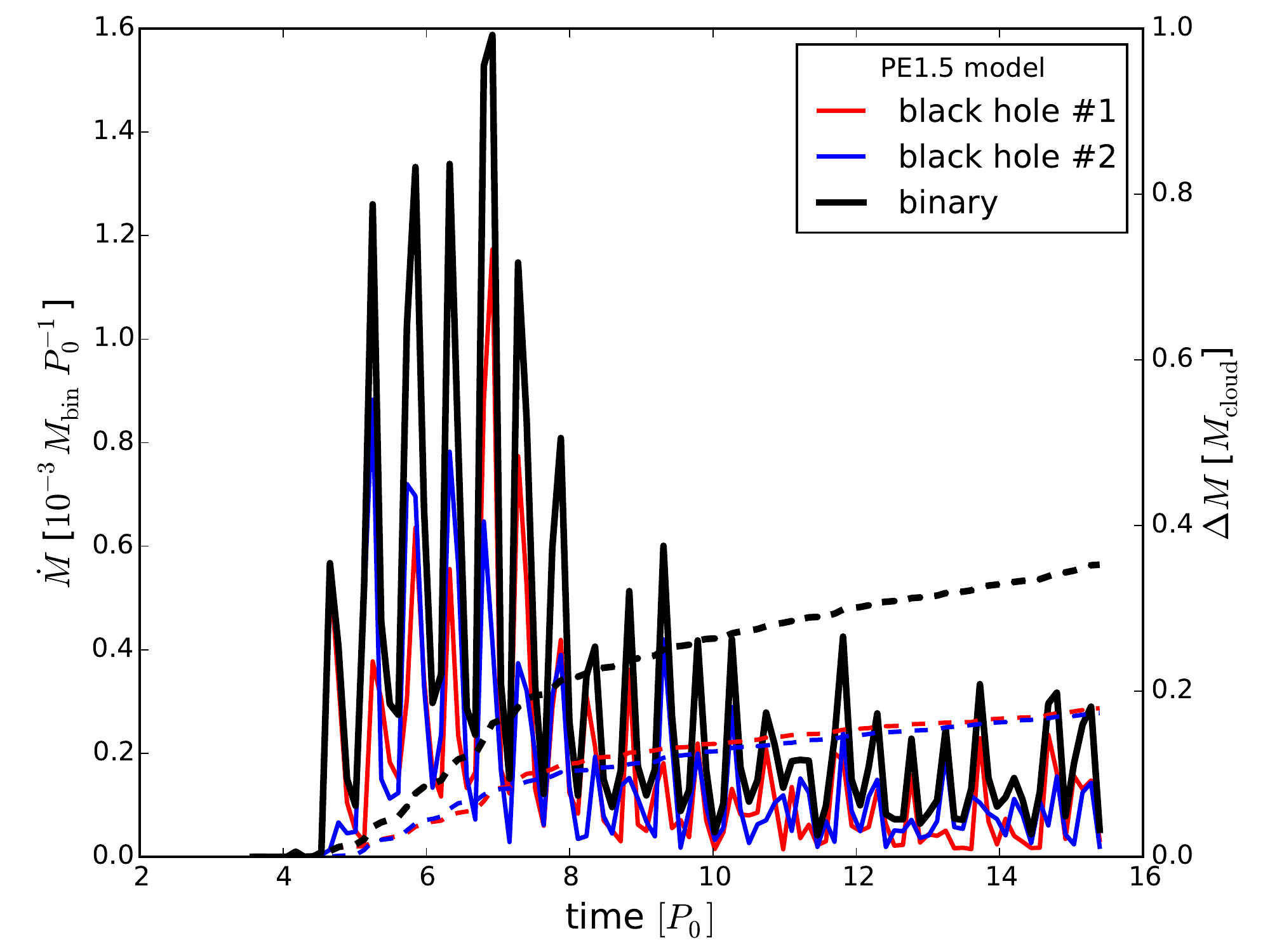}}
  \put(333,130){\includegraphics[width=0.3333\textwidth]{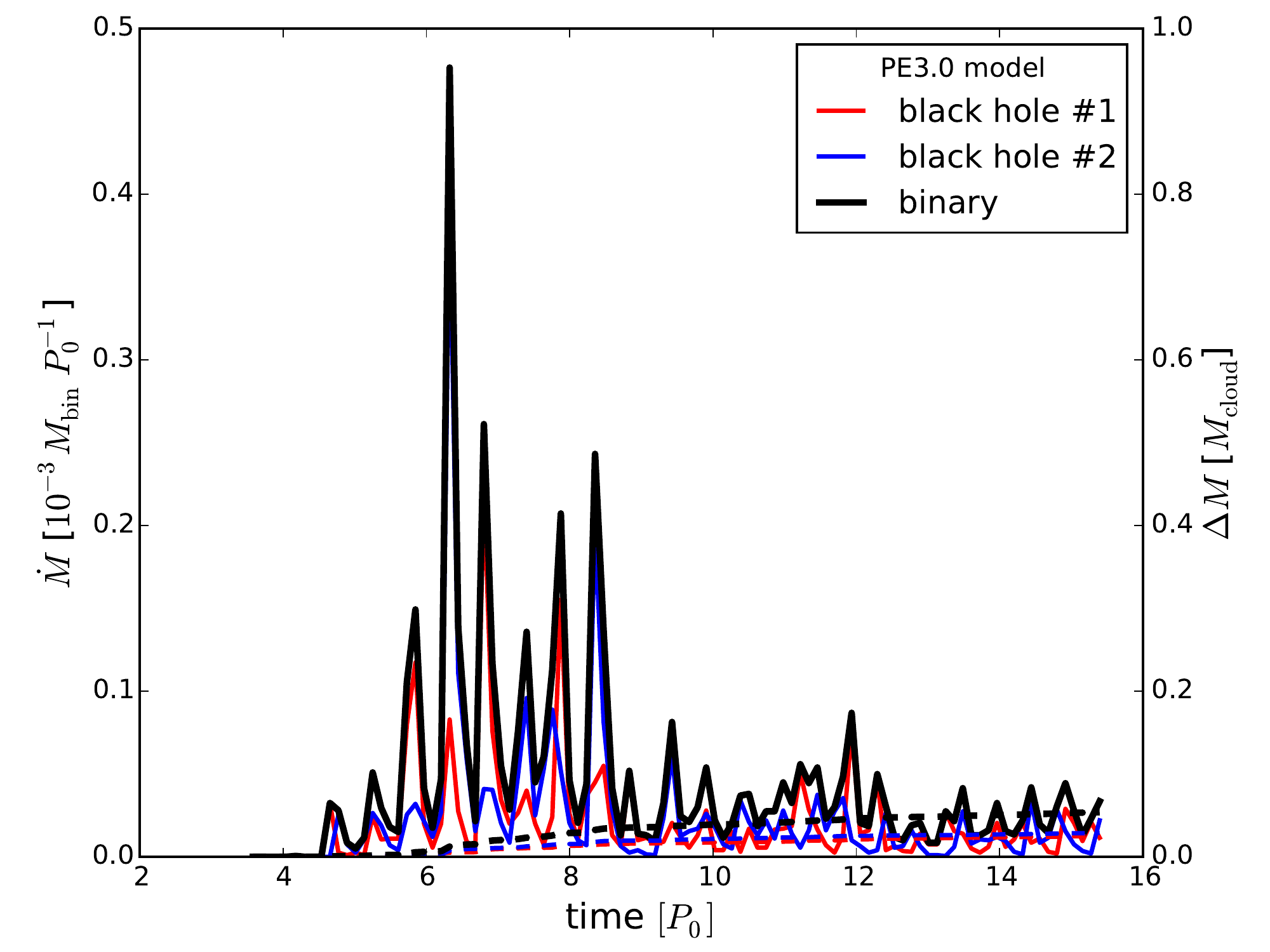}}
  \put(0,260){\includegraphics[width=0.3333\textwidth]{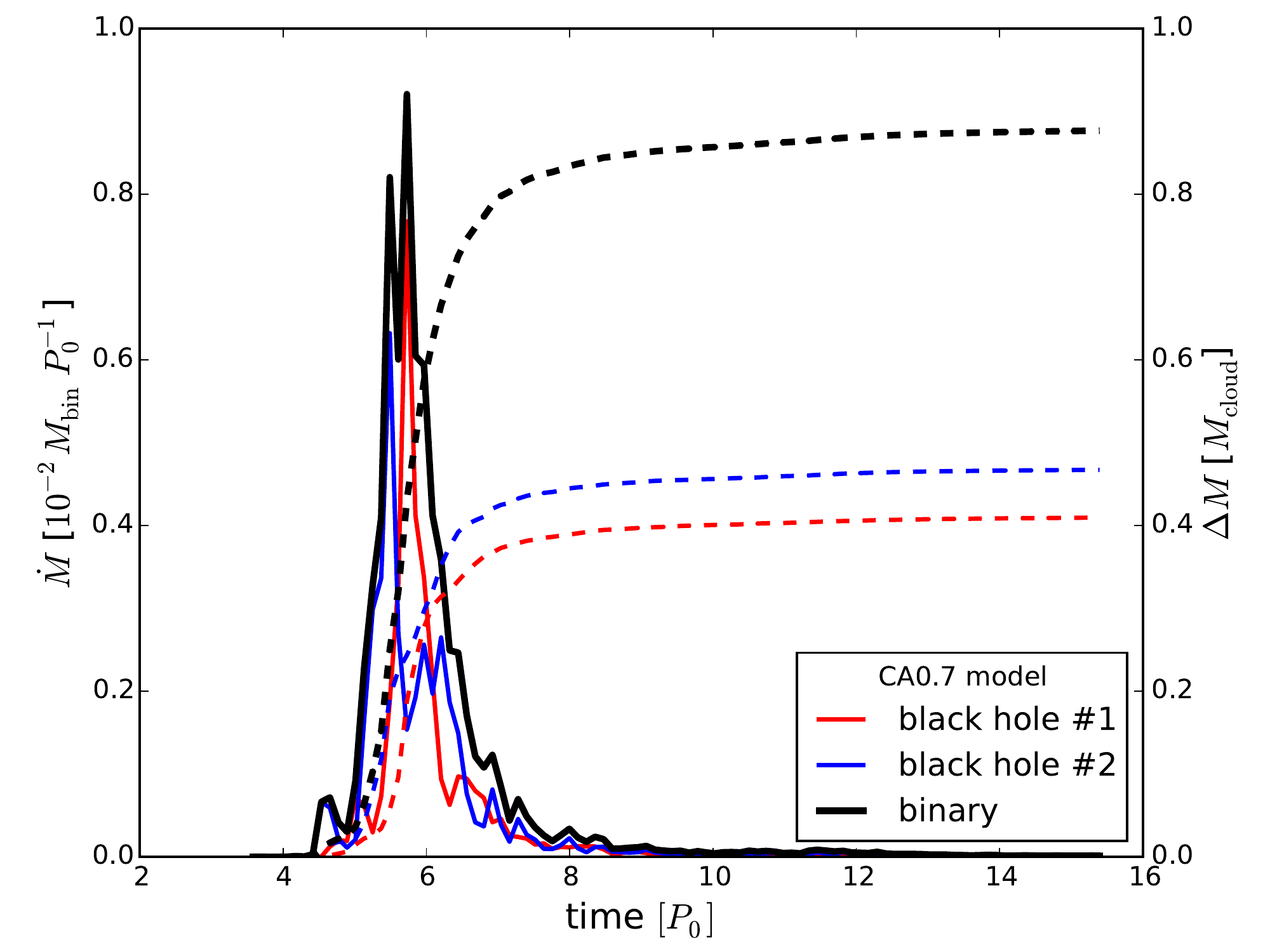}}
  \put(167,260){\includegraphics[width=0.3333\textwidth]{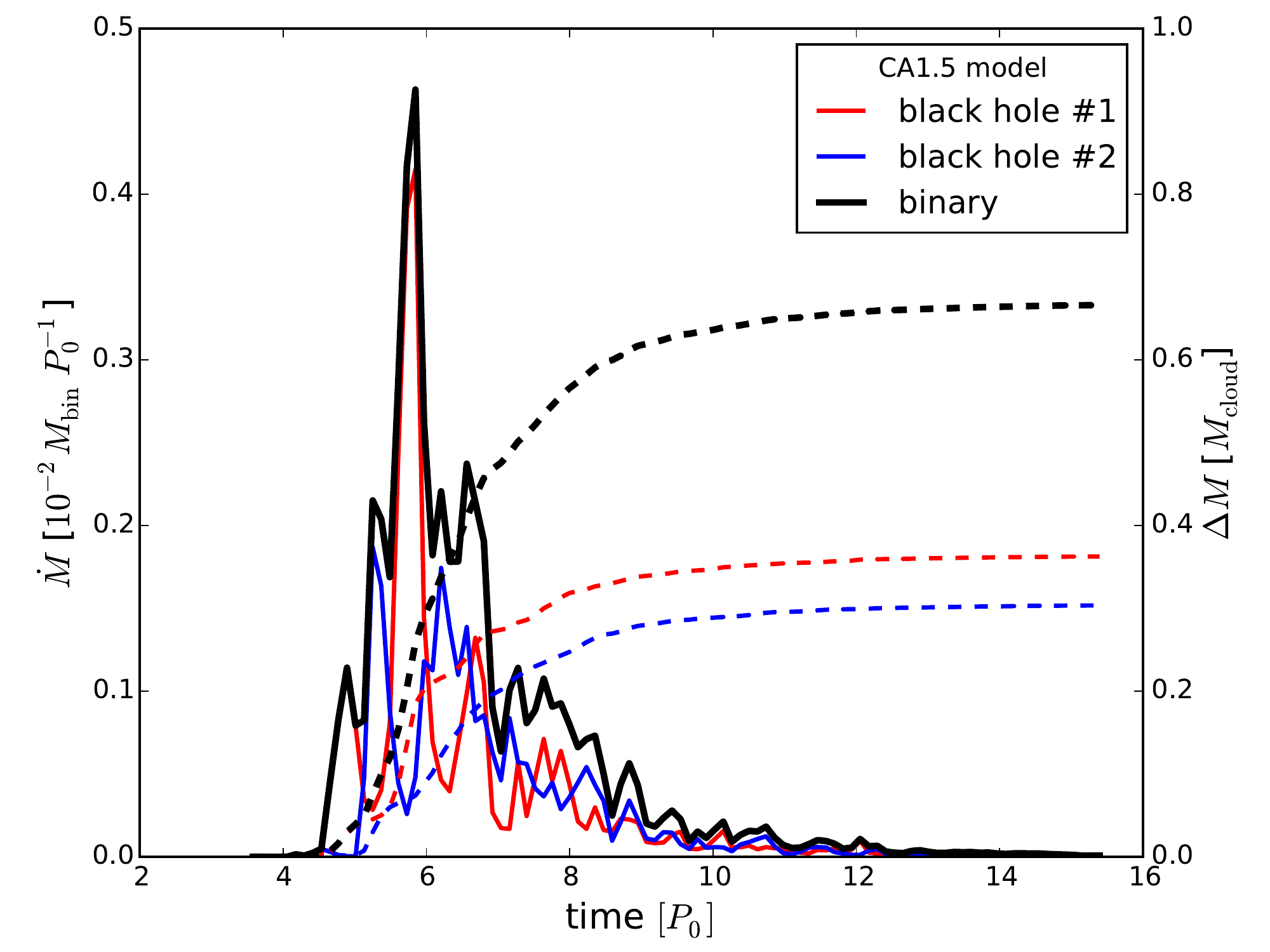}}
  \put(333,260){\includegraphics[width=0.3333\textwidth]{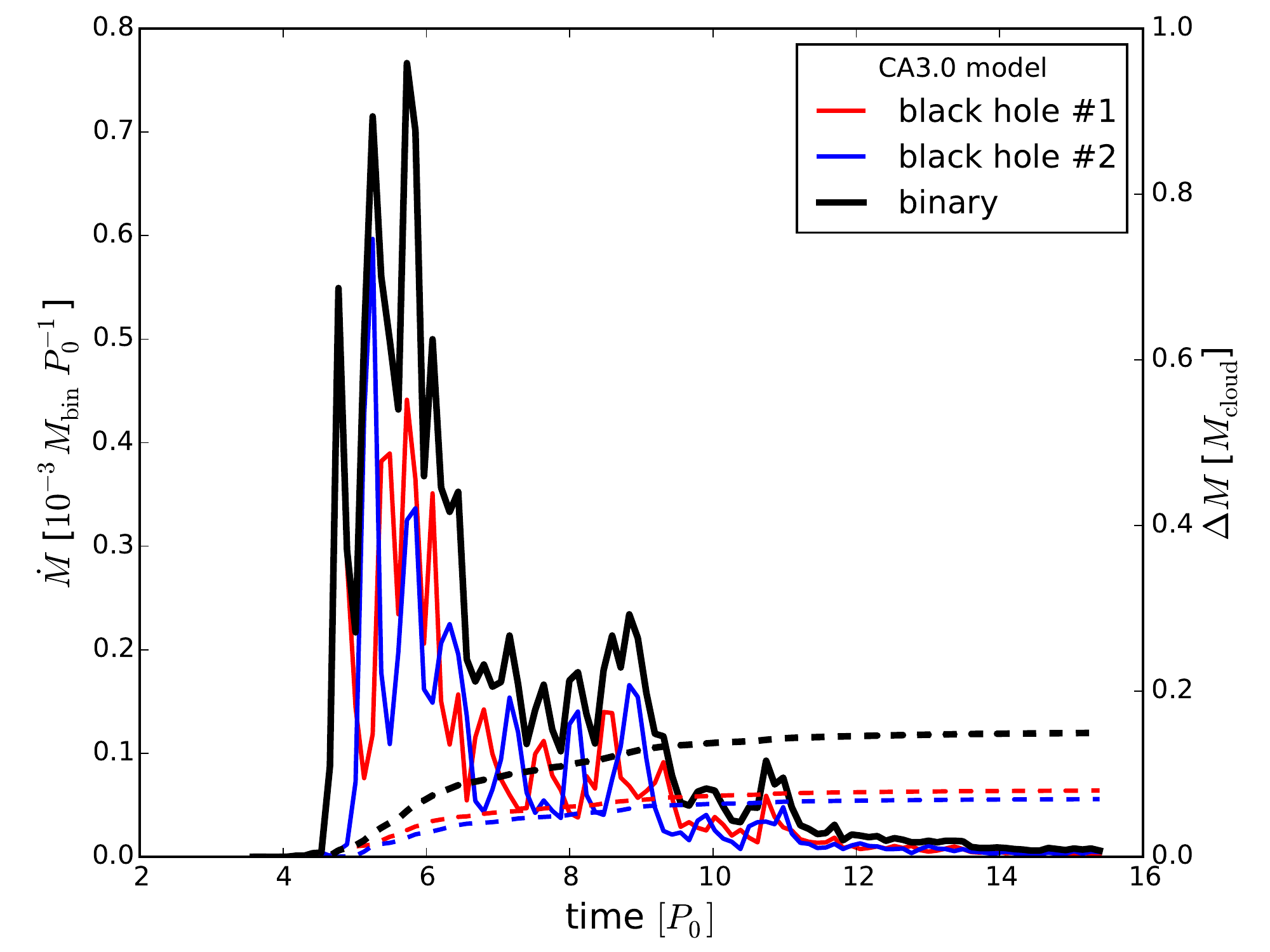}}
  \put(0,390){\includegraphics[width=0.3333\textwidth]{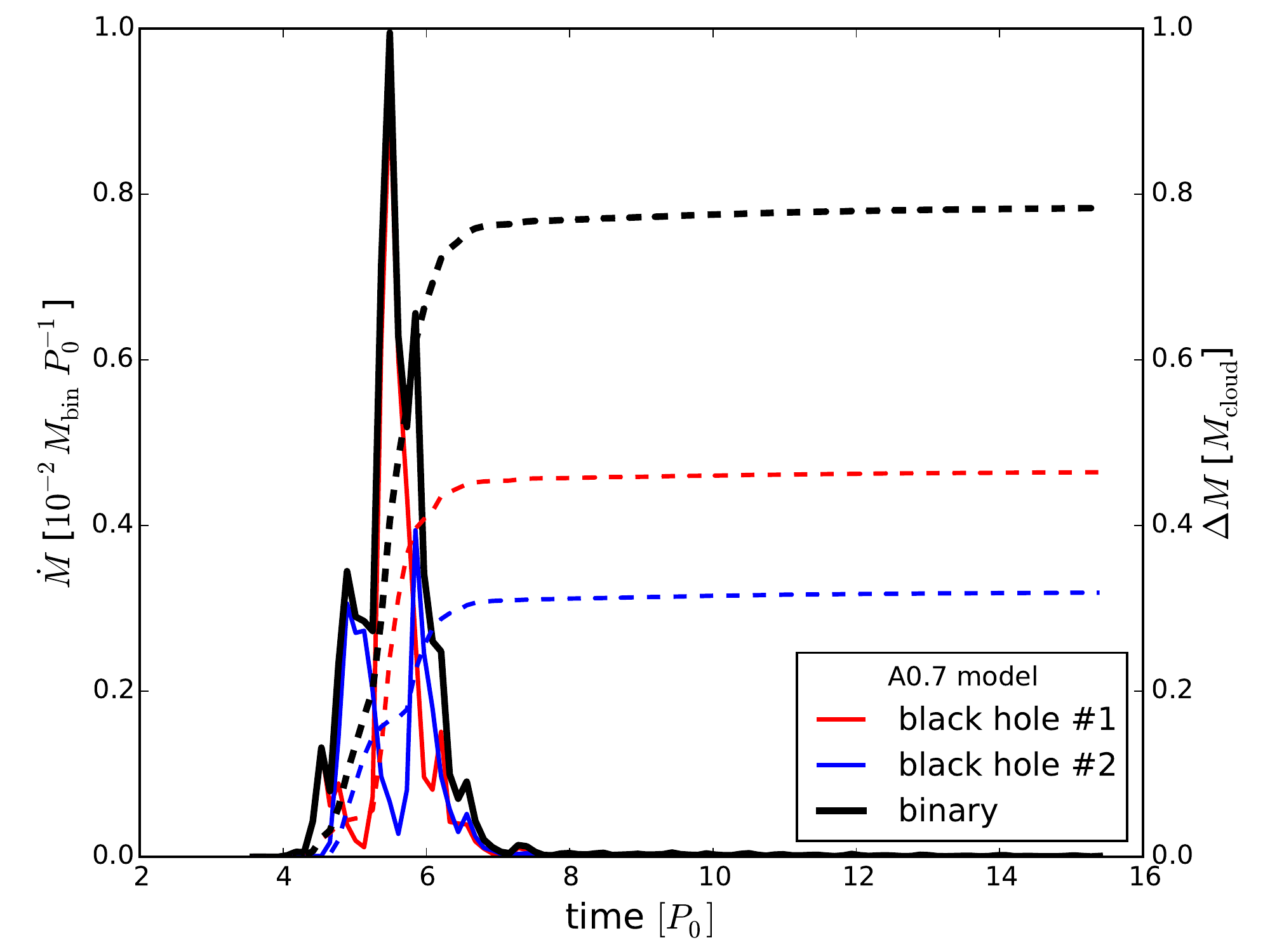}}
  \put(167,390){\includegraphics[width=0.3333\textwidth]{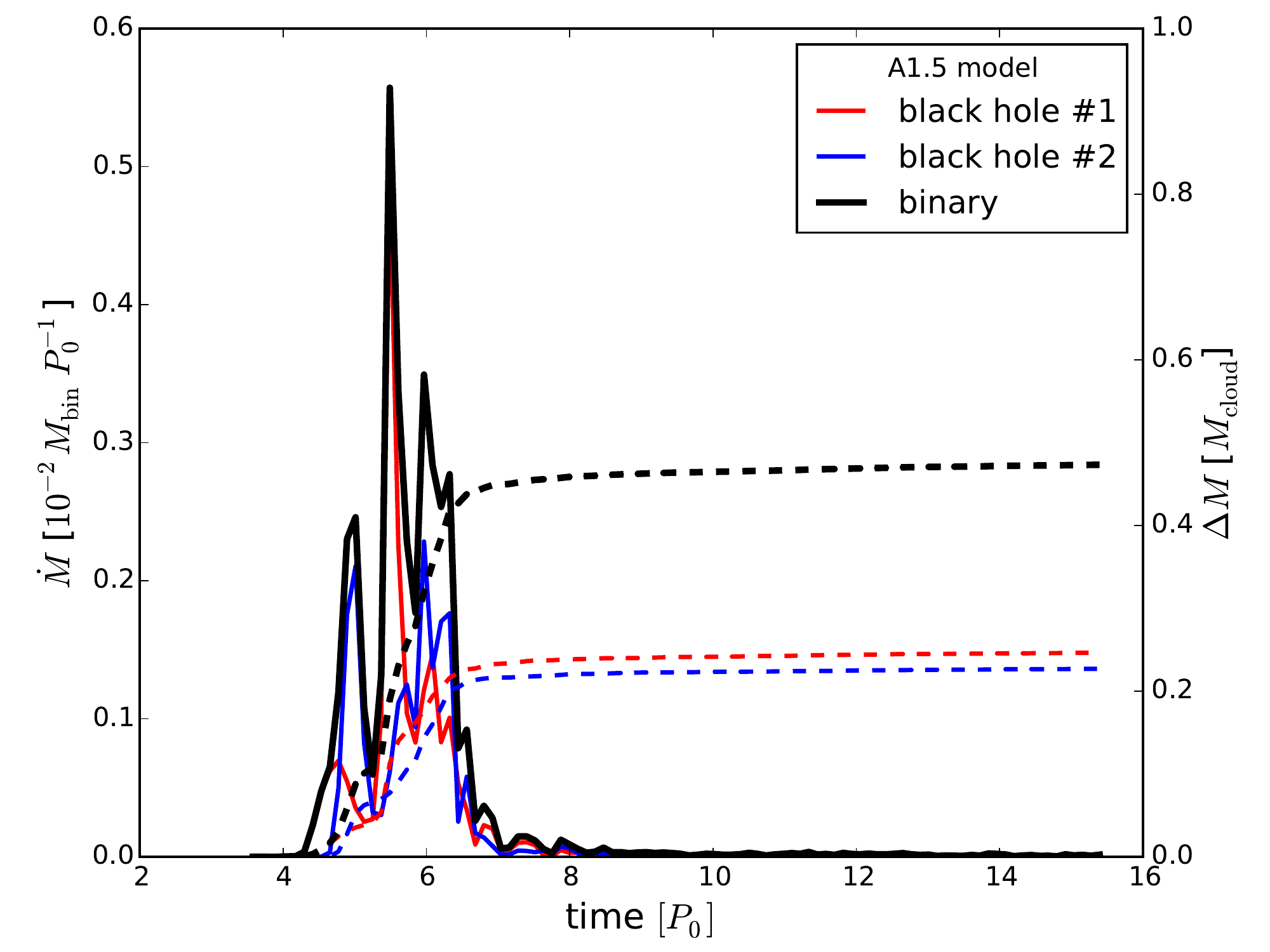}}
  \put(333,390){\includegraphics[width=0.3333\textwidth]{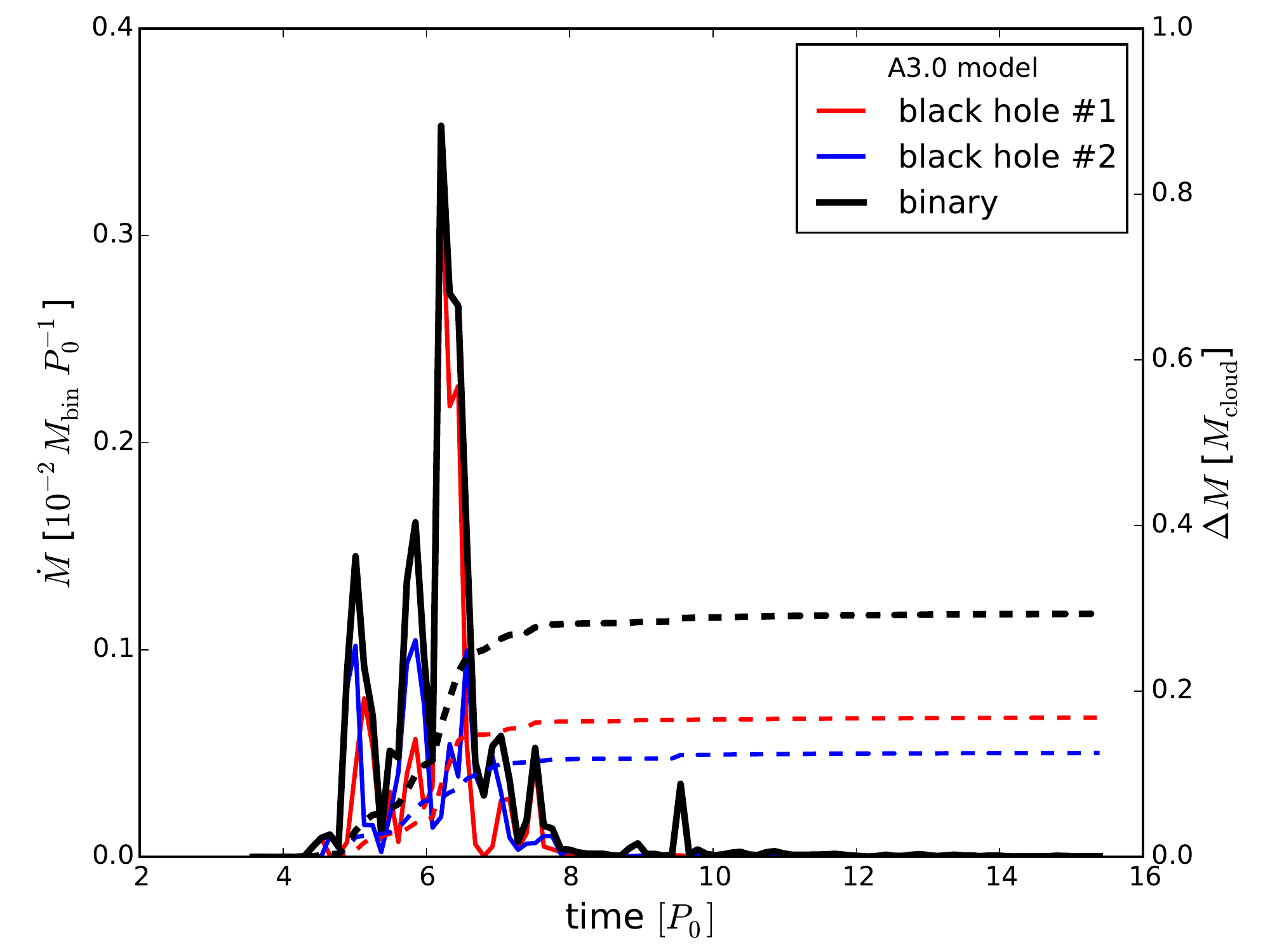}}
 \end{picture}
\caption{Evolution of the accretion rate (solid, left axes) and cumulative accreted mass (dashed, right axes) onto the binary (black) and each black hole (blue and red), for all simulations. From left to right: pericentre distances of 0.7, 1.5 and 3$R_{\rm bin}$. From top to bottom: aligned, counter-aligned, perpendicular edge-on and perpendicular face-on orbits. Notice that the range  of the left y-axes is not the same for all plots. On each curve we see a periodic behaviour that is directly related with the binary period, 
and in most cases the accretion on to both SMBHs are in counter-phase with each other.}
\label{accretion}
\end{figure*}

 In Fig.~\ref{accretion} we show the accretion rate and cumulative accreted mass for every inclination and impact parameter. Note that throughout this section the results  are presented in code units, thus we can re-scale the results to a range of binary masses and periods (for physical rescaling of our results see Section \ref{sec_fin}).

It is important to emphasise that the accretion radius set for each SMBH is very large compared to the Schwarzschild radius, so the accretion rate that we compute does not correspond to actual accretion onto the SMBHs. The gas within our accretion radius will have non-zero angular momentum, thus should settle on an accretion disc and evolve on a viscous timescale ($t_{\rm visc}$), which is typically longer than its dynamical timescale ($t_{\rm dyn}$) in the SMBH gravitational potencial. In consequence, our accretion rates correspond to the rate at which the gas is added to the BH-accretion disc systems, which are unresolved in our simulations. For an $\alpha$-disc \citep{SS1973}, these timescales are related as following:
\begin{equation}
t_{\rm visc}\sim \frac{1}{\alpha}\left(\frac{H}{R}\right)^{-2}t_{\rm dyn},
\end{equation}
where $\alpha$ is a dimensionless number quantifying the strength of the viscosity (with typical values of $\sim 0.01-0.1$) and $H/R$ is the disc aspect ratio. On the other hand, the dynamical timescale (at the sink radius) can be related with the orbital period of the binary as following:
\begin{equation}
\frac{t_{\rm dyn}}{P_{\rm bin}}\sim \frac{1}{2\pi}\left(\frac{r_{\rm sink}}{a}\right)^{3/2}\left(\frac{M}{M_1}\right)^{1/2} \approx 5\times 10^{-3},
\end{equation}
where $r_{\rm sink}=0.1a$ is the accretion radius, $M$ is the binary mass and $M_1=0.5M$ is the mass  of one SMBH.

As a rough estimate, we would expect the actual accretion rate onto each SMBH be related with the binary orbital period if the latter is longer than the viscous timescale. In other words, when
\begin{equation}
\alpha\left(\frac{H}{R}\right)^{2}\gtrsim 5\times 10^{-3},
\end{equation}
which implies that there are combinations of parameters (e.g., $\alpha\sim0.1$ and $H/R\sim0.2$) for the unresolved accretion discs where the variability presented below would indeed affect directly the actual accretion on to the SMBHs \citep{Sesana2012}. For systems where the viscous time is longer than the binary orbital period, the streaming periodicity will not be directly reflected into a variable accretion onto each SMBH, and a pair of persistent (unresolved) mini-discs will form.  However, even in these cases we could expect to find observational signatures of this variability (\S~\ref{sec_fin}).

Another caveat of our model is that we do not include any type of feedback from the SMBHs; radiation pressure could reduce the amount of gas that is accreted. Nevertheless, the amount of gas that actually reaches the SMBHs will likely be a monotonic function of the value that crosses our sink radius.

In summary, the behaviour of the accretion rate shown in this section can still be useful to characterise the accretion onto the SMBHs and the mini-discs emission.

As expected, the accretion rate is very high at the beginning of the interaction with the cloud.  Most of the cloud is engulfed by the binary during the first few orbits ($\sim 4-7$), as we can observe in the cumulative mass of each figure. This stage corresponds to the first passage of the cloud. After that interaction the accretion drops sharply, by 2 or 3 orders of magnitude in some cases. 

For most orbital configurations there is a clear periodicity on the accretion rates. Each black hole has an accretion periodicity which matches the binary orbital period, but in phase opposition with each other, so that the total binary accretion rate has a period of half an orbit. This feature is associated with the stream of gas that remains bound to the binary and feeds each black hole alternatively when they cross the stream. Variability related to the binary orbital period is the type of behaviour that we expect will help to identify and characterise these systems.

The accretion in the cases of perpendicular orbits (edge and face-on, two lower rows of Fig.~\ref{accretion}) tends to be more extended than the parallel ones (two upper rows), in the sense that there are still significant peaks after the first passage of the cloud. This is because, as explained in the previous section, the slingshot is less efficient when the encounters have perpendicular relative velocities, allowing more material to remain around the binary in close orbits. 

For all the orbital configurations we model, we obtain two trends with increasing impact parameter: (i) the accretion rates and the total accreted mass decrease, and (ii)   the relative amplitude of the accretion rate peaks during the cloud first passage decreases compared to the later stages.

\begin{figure}
\centering
\includegraphics[width=0.46\textwidth]{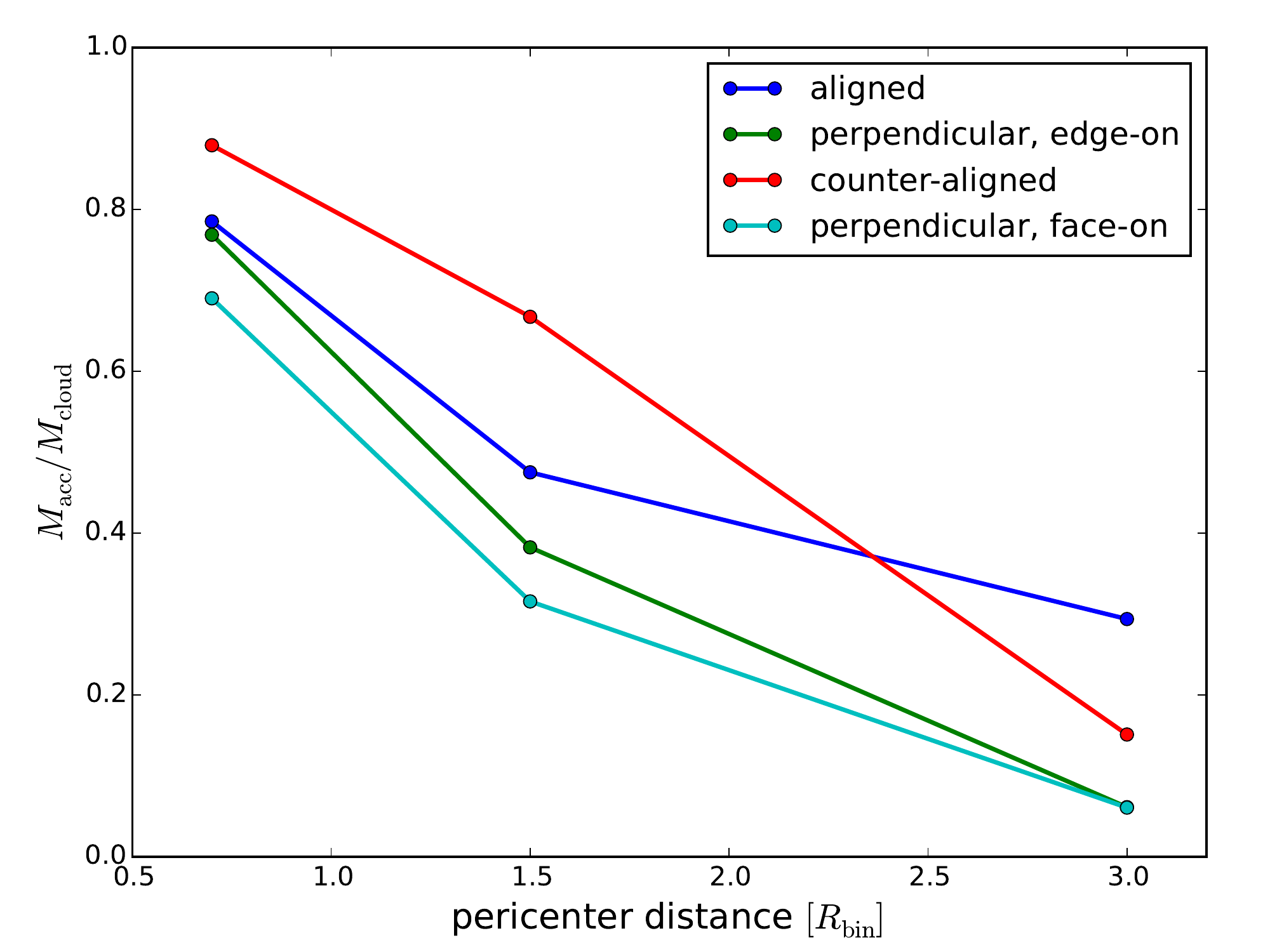}
\caption{Fraction of the total cloud mass accreted by the binary as a function of the pericentre distance, for the four different orbital orientations. Notice that the total mass is strongly dependent on the impact parameter of the binary, dropping by one order of magnitud in some cases.}
\label{acc_b}
\end{figure}

The compilation of the total mass accreted by the binary at the end of each simulation is shown in Fig.~\ref{acc_b}. Here we can see how the accretion is dramatically reduced when we increase the impact parameter of the cloud. For instance, in the PE3.0 model the total accretion is around 5\% of the cloud mass, in contrast with the $\approx$ 70\% on the smaller pericentre distance (PE0.7). This is very interesting as it shows the transition between a ``prompt accretion" regime, and one characterised by  the formation of circumbinary discs for the aligned and perpendicular edge-on orbits (see Section \ref{sec_cbd}).  

\section{Misaligned mini-discs}
\label{sec_mds}

 As we mention in Section \ref{sec_discs}, the only configurations that show the formation of extended and stable mini-discs, limited by the Hills radius, are the aligned models, which is the first case we study below. However, we also measure the direction of the gas around the SMBHs for the other inclinations to see if possible unresolved mini-discs have some preferential orientation.

\subsection{Mini-discs for the aligned configurations}

\begin{figure*}
\centering
 \begin{picture}(500,280)
  \put(0,140){\includegraphics[width=0.5\textwidth]{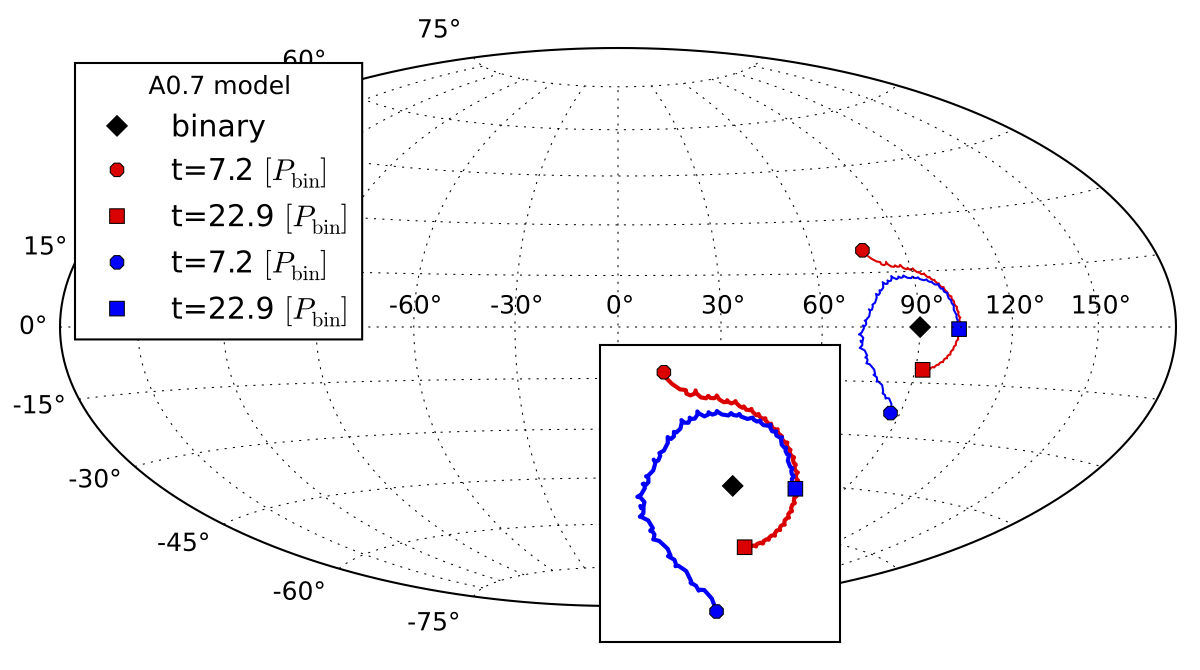}}
  \put(250,140){\includegraphics[width=0.5\textwidth]{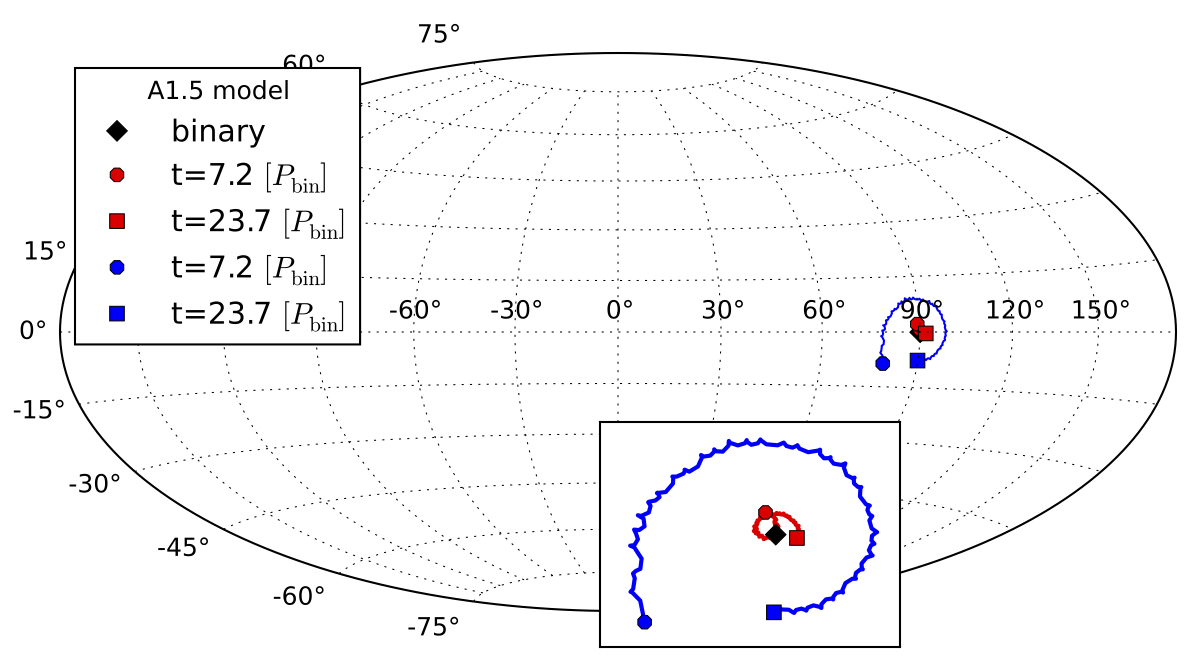}}
  \put(120,0){\includegraphics[width=0.5\textwidth]{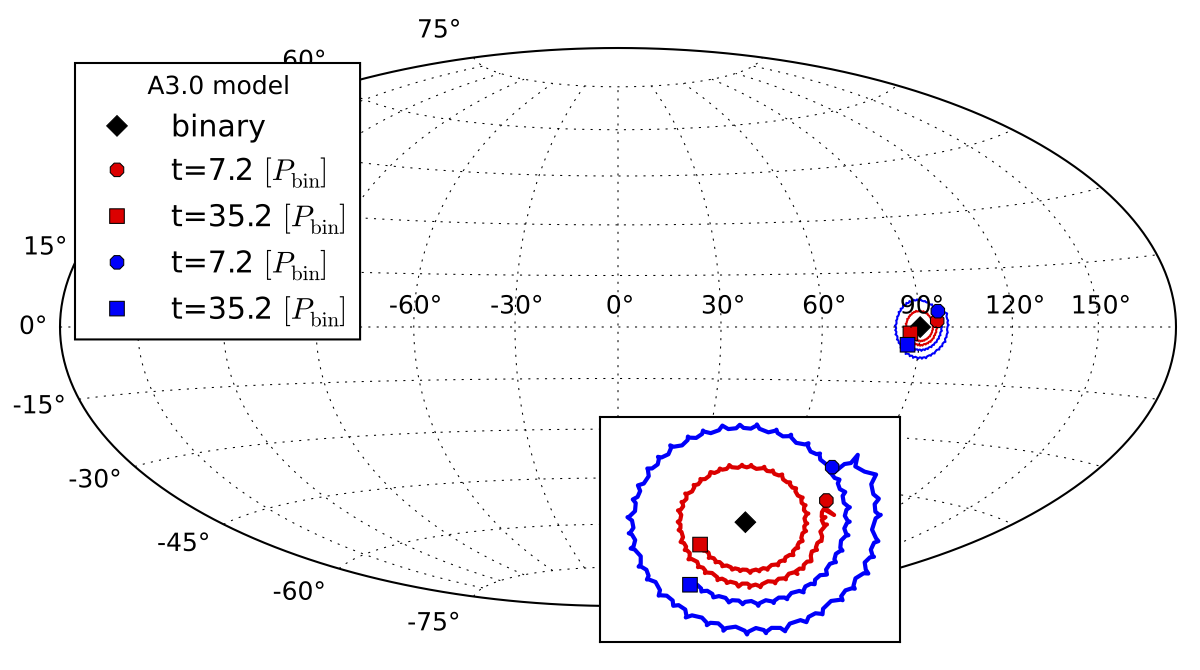}}
 \end{picture}
\caption{Time evolution of the directions of both mini-discs shown in Hammer projections for the aligned orbits and the three different impact parameters. Upper left: $0.7R_{\rm bin}$, upper right: $1.5R_{\rm bin}$ and bottom: $3R_{\rm bin}$. The circles and squares correspond to the times when each mini-disc appears and the end of the simulation, respectively; while the black diamonds correspond to the binary orbit orientation. The inset panels show zoom-ins of the projections, in which is noticeable that the evolution of the mini-discs show two combined effects: a steady precession around the aligned position and a super-imposed wobbling.}
\label{mds_alig}
\end{figure*}

In all our models with aligned orbits we see the formation of prominent and persistent mini-discs around each SMBH (top row of Fig.~\ref{BHB}). In order to measure the level of alignment with the orbit of the binary we compute the mini-disc direction using the total angular momentum vector of the gas particles within the Hill sphere. We represent this direction with the spherical coordinates $(\theta,\phi)$ in the reference frame of the corresponding SMBH. The time evolution of the mini-disc directions is shown in the Hammer projections of Fig.~\ref{mds_alig}.

From these projections we observe that the mini-discs are well defined and evolve smoothly with time, and also that they are roughly aligned with the binary orbit (and the original cloud orbit), although there is always some level of misalignment.  Studying closely the time evolution of the minidisc orientations, we notice the following behaviour:  not only the mini-discs precess around the aligned position, but on top of that we observe another low-amplitude, periodic motion that hereafter we  refer to as ``wobbling". In order to study these motions in more detail, we describe the direction of the mini-discs with other two angles: inclination and position. The first one is the angular difference with respect to the aligned position, while the second is the angle between the projection of the mini-disc on the orbital plane and an arbitrary vector on the same plane. 

\begin{figure}
\centering
\includegraphics[width=0.46\textwidth]{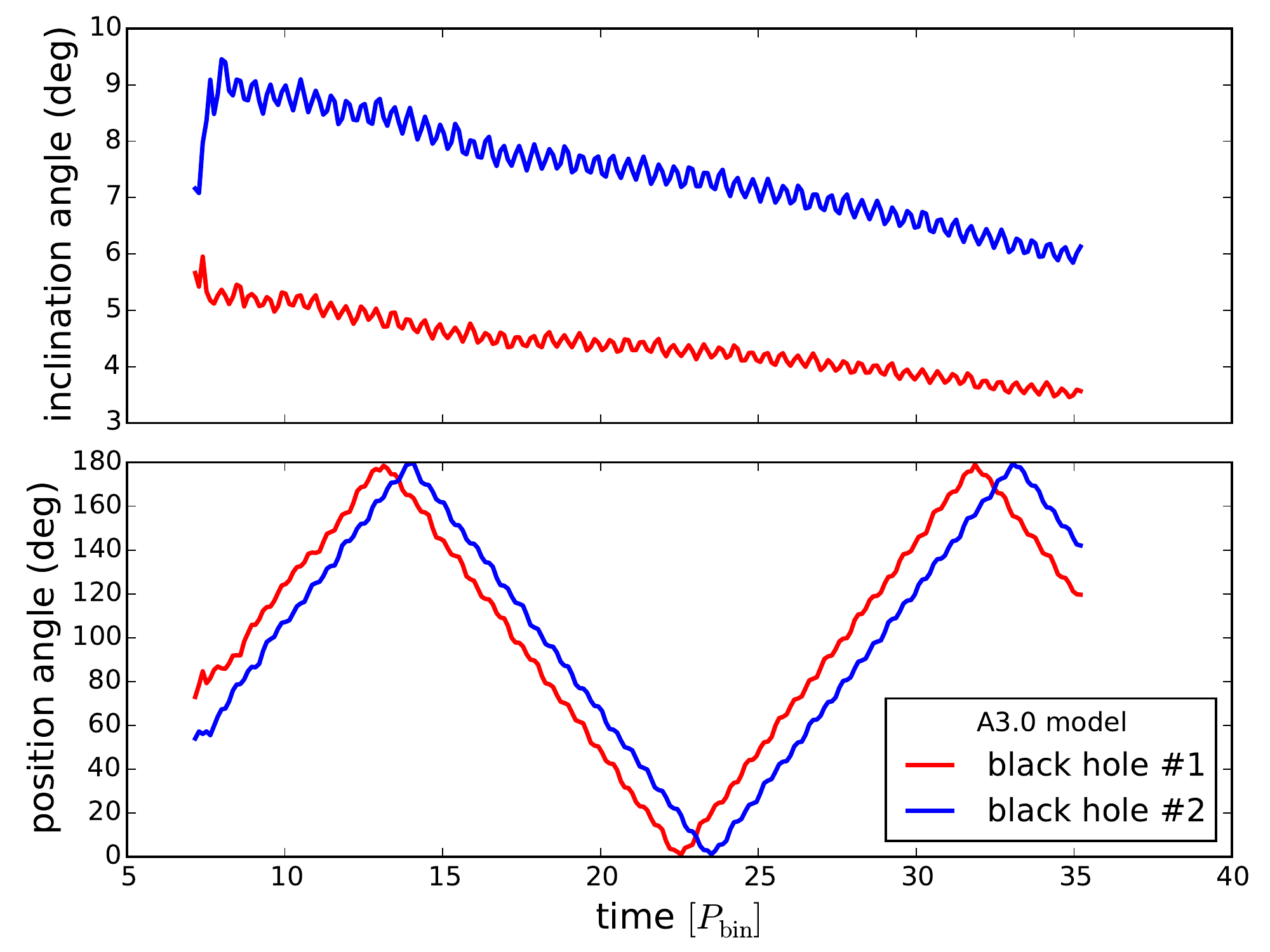}
\caption{Upper panel: inclination angle evolution of the mini-discs respect to the aligned position. Both the slow decline due to dissipation and the wobbling discussed in the main text are clearly visible.
  Lower panel: position angle evolution, i.e. angle between the projection of the mini-disc direction on the orbital plane and an arbitrary vector on the same plane. This evolution shows how the discs slowly and steadily precess around the aligned position. The period of this movement is around 20 binary periods.} 
\label{precess}
\end{figure}

Since a similar behaviour is observed in all three runs with aligned orbits (see  Fig.~\ref{mds_alig}), 
we now concentrate our analysis in the A3.0 model, which ran the longest and has the better defined direction evolution.  
Notice that this run also shows a slow but clear alignment of the mini-discs with the binary orientation.  However, we do not discuss this in detail as it depends on the numerics (see below).   
We show the time evolution of the inclination and position angles in Fig.~\ref{precess}. We observe that the precession is steady (constant slope in absolute value), with a period of around 20 binary orbits. On the other hand, the mini-discs inclination tend to decrease with time and to wobble with a period almost exactly half of the binary period. As we show below, this behaviour is expected for misaligned discs.

\subsection{Dynamics of misaligned discs}
Based on the study by \citet{Bate2000} we can obtain an analytical understanding of the dynamics of our misaligned mini-discs around each SMBH. We consider a circular binary with components $M_1$ and $M_2$, with separation of $a$. We sit on a non-rotating frame with $z$ axis parallel to the rotation axis of the binary, centred on one of the SMBHs. The force at a position vector $\vec r$ (with $r\ll a$) caused by the other black hole is given by:
\begin{equation}
\vec F_2=-\frac{GM_2m_r}{a^3}\vec r+\frac{3GM_2m_r}{a^5}(\vec r\cdot\vec a)\vec a,
\label{force}
\end{equation}
where $\vec a$ is the position vector of the secondary SMBH and $m_r$ is a test mass.

The right side of equation \eqref{force} can be split into 2 contributions: the isotropic term ($m=0$) and the quadrupole ($m=2$), and each of them can be associated with a different effect.

\begin{enumerate}
\item The $m=0$:
If we consider a ring of gas with mass $m_r$ and radius $a_r$, the torque produced by this term is:
\begin{equation}
\vec T_0=-\frac{3}{4}\left(\frac{GM_2m_ra_r^2}{a^3}\right)\sin\delta\cos\delta \hat\imath.
\label{T0}
\end{equation}

The angular frequency of a ring of disc material is given by
\begin{equation}
\vec\Omega_d=-\Omega_d\sin\delta\hat\jmath + \Omega_d\cos\delta\hat k,
\label{}
\end{equation}
where $\Omega_d$ is a function of the radius of the considered ring, $\delta$ is the inclination angle of the ring respect to the binary plane. Then, as $\vec\Omega_d\cdot \vec T_0=0$, the net effect of this torque is to produce a precession around the $z$ axis. The mean precession rate is
\begin{equation}
\frac{\omega_p}{\Omega_d}=\frac{3}{4}q\cos\delta\left(\frac{a_r}{a}\right)^3,
\end{equation}
where $q$ is the mass ratio of the binary.
\item The $m=2$:
In this case the torque on the ring is
\begin{equation}
\begin{split}
\vec T_2=-\frac{3}{4}\left(\frac{GM_2m_ra_r^2}{a^3}\right)\sin\delta\left[\cos\delta\cos(2\Omega_b t) \hat\imath\right.\\
+\cos\delta\sin(2\Omega_b t) \hat\jmath+\sin\delta\sin(2\Omega_b t)\hat k],
\end{split}
\label{T2}
\end{equation}
where $\Omega_b$ is the binary angular frequency. This oscillating torque is also perpendicular to the rotation of the ring $(\vec T_2\cdot\vec \Omega_d=0)$, and the effect is to produce an oscillation around the steady precession, with a frequency equal to twice the binary frequency, consistent with what we measure for the inclination in Fig.~\ref{precess}. The amplitude of the wobble is roughly $\sim \omega_p/(2\Omega_b)$.
\end{enumerate}

The behaviour of the entire disc will be given by the integral of each ring. Then, the net precession rate is
\begin{equation}
\frac{\omega_p}{\Omega_d}= K\cos\delta q \left(\frac{R}{a}\right)^3,
\end{equation}
where 
\begin{equation}
K=\frac{3}{4}R^{-3/2}\frac{\int_0^R\Sigma r dr}{\int_0^R\Sigma r^{-3/2} dr},
\end{equation}
and $\Sigma$ is the mass surface density of the disc. Some typical values of $K$ are: $K=15/32\approx 0.487$ for a constant surface density profile \citep{Larwood1996}; and $K=3/10$ for $\Sigma\propto r^{-3/2}$ \citep{Hartmann1998}.

Thus typically
\begin{equation}
\frac{\omega_p}{\Omega_b}\approx 0.05\left(\frac{K}{0.04}\right)q\sqrt{\frac{2}{1+q}}\left(\frac{R}{0.3a}\right)^{3/2}\cos\delta
\label{omega_p}
\end{equation}

Evaluating equation \eqref{omega_p} with the approximate values of our mini-discs we find
\begin{equation}
\frac{P_p}{P_b}\approx 20,
\end{equation}
and also the amplitude of the wobble in the outer part of the disc, which is the dominant contribution, is
\begin{equation}
A_{\rm wob}\sim\frac{\omega_p}{2\Omega_b}\approx 0.02\;\mbox{rad}\approx 1\;\mbox{deg}.
\end{equation}
Both quantities are actually very close to what we observe on Fig.~\ref{precess}.

All these calculations are made assuming that the discs are able to communicate the precession efficiently without breaking, which is clearly the case of our simulations as we observe them moving as whole. 
However, the ability of the disc to precess rigidly depends strongly on its aspect ratio and viscosity \citep{Pap1995,Larwood1996,Lubow2000,Fragner2010,Dogan2015}, which are not well resolved quantities in our models due to the small number of particles ($\sim 1000$) that shape our mini-discs and the numerics itself. In addition, the presence of the accretion radius around the sink particles excises the inner portion of the mini-discs, which does not allow us to model the precession and wobbling particularly onto those scales.

The numerical models performed by \citet{Fragner2010}, study the evolution of individual discs that arise in misaligned binary systems, showing precession and wobbling (i.e. periodic perturbation in the inclination angle), similarly as we found in our mini-discs. Their grid-based hydrodynamic code allows them to control better the viscosity's influence in the disc evolution. They show that, for several combinations of aspect ratios and viscosities, the discs will efficiently communicate the differential precession and move as a rigid body. In particular, they find that thin discs ($h\lesssim 0.03$) with high viscosities will achieve a state of rigid precession after developing a twist. More important, when the discs are not disrupted due to the differential precession, the periodic perturbations in the inclination are able to travel all the way to the central parts (cf. their Fig.6), which we are not able to resolve.

In conclusion, even if we are not able to robustly study the global dynamics of the mini-discs due to the relatively low resolution we can afford with our models, we expect that the misalignment arising from the infall of extended portions of gas will produce the precession and wobbling we observe.

\subsection{Mini-discs for other inclinations}

\begin{figure*}
\centering
 \begin{picture}(500,280)
  \put(0,140){\includegraphics[width=0.5\textwidth]{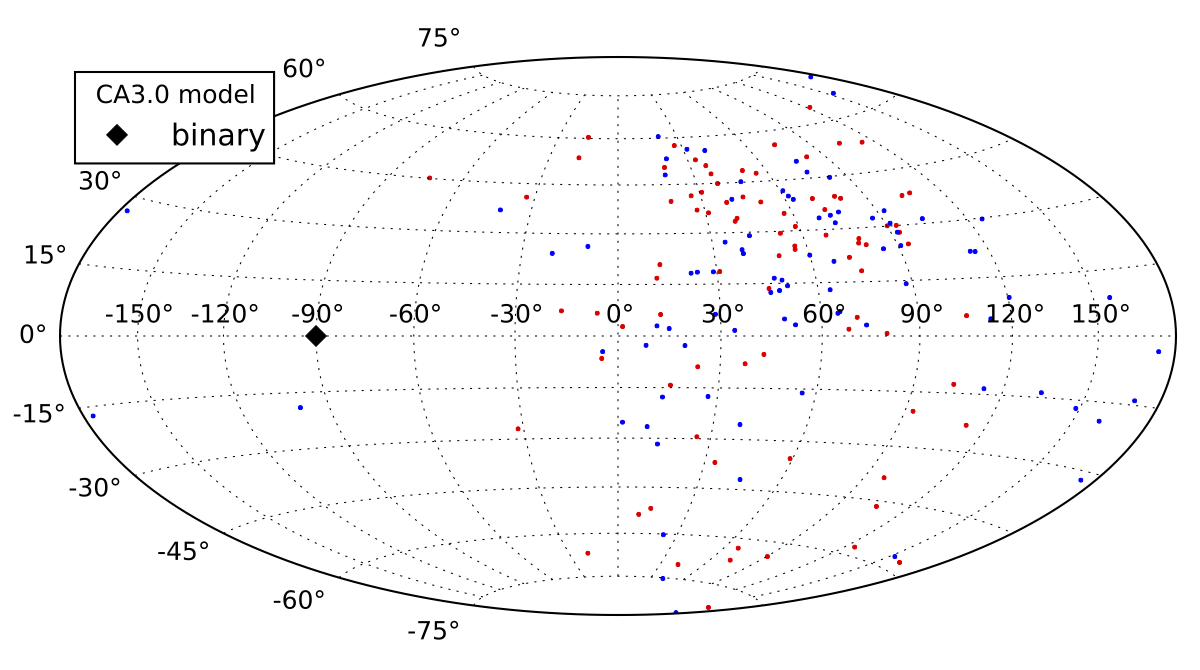}}
  \put(250,140){\includegraphics[width=0.5\textwidth]{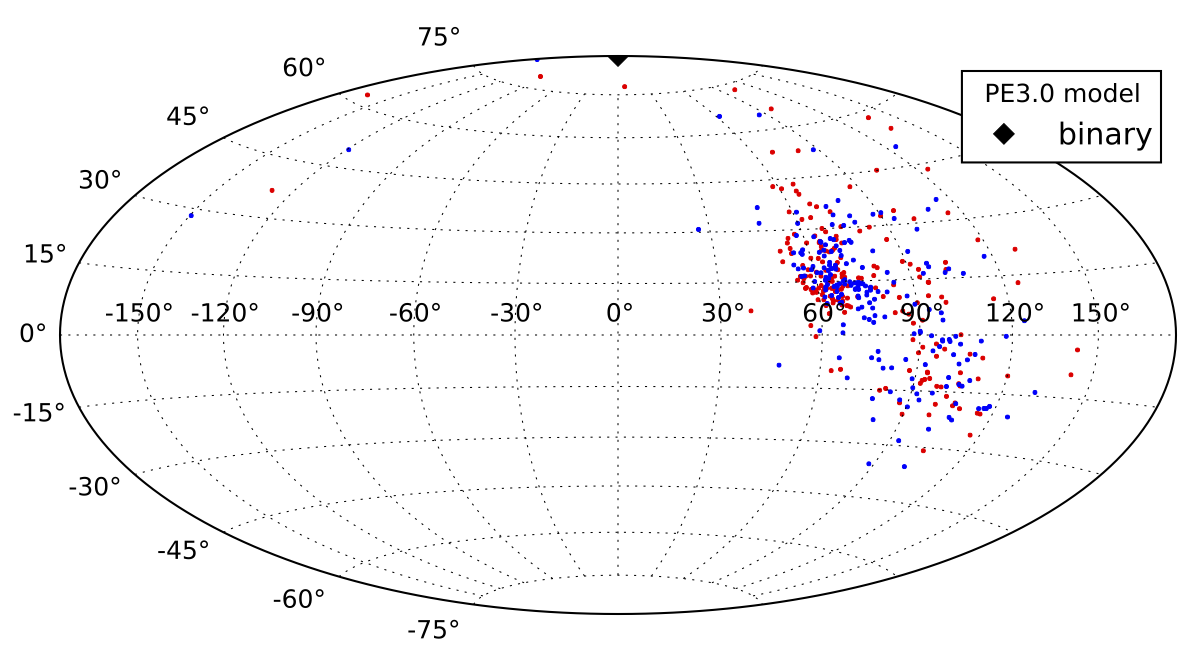}}
  \put(120,0){\includegraphics[width=0.5\textwidth]{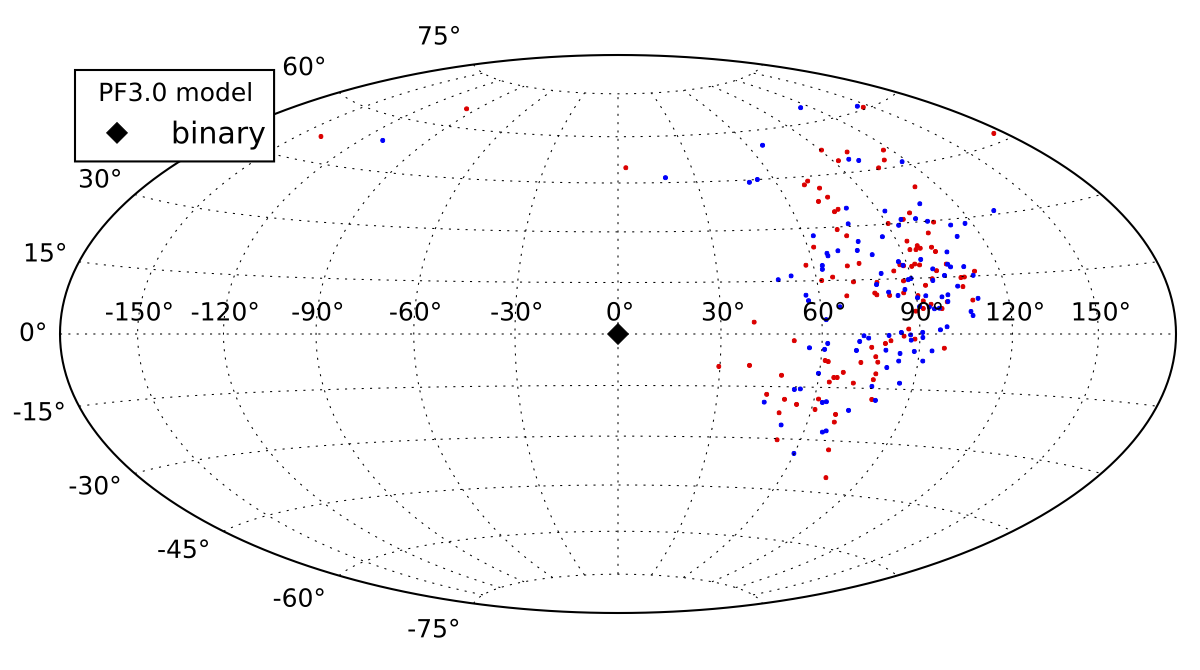}}
 \end{picture}
\caption{Evolution of the direction of all the accreted gas by each SMBH (red and blue dots) in a Hammer projection. The different plots correspond to different inclinations: counter-aligned (upper left), perpendicular edge-on (upper right) and perpendicular face-on (bottom). As indicated in the legend, all correspond to the larger impact parameter. The black diamonds indicate the binary direction.  Notice how in all cases the gas tends to follow the initial orientation of the cloud, located always at $(90^\circ,0^\circ)$, and not the orientation of the binary.  The effect is weakest for the counter-aligned orbit case.}
\label{mds_other}
\end{figure*}

As explained in Section~\ref{sec_discs}, during the whole duration of the simulations with different inclinations (counter-aligned and perpendiculars) we do not observe the formation of persistent mini-discs, probably because we do not have the spatial resolution due to the artificially large accretion radii and the low angular momentum of the captured gas relative to the individual SMBHs.
Nevertheless, we can still measure whether the material around each SMBH has some preferential direction. We do so by computing the angular momentum of all accreted gas between outputs, which is an indicator of the direction that the unresolved mini-discs might have, if they exist. 
 We consider the time after the first passage of the cloud ($\sim8$ orbits) for every snapshot until the simulation ends. We show an example of the projections obtained with the larger impact parameter in Fig.~\ref{mds_other}, where each point represents the direction of the gas in a particular output.

For the counter-aligned orbits (upper left panel of Fig.~\ref{mds_other}) the different orientations of the gas are not concentrated around any particular point. However, they are preferentially on the right side of the projection, which indicates that the gas tends to be counter-aligned respect to the binary rotation.
On the other hand, for both perpendicular orbits (upper right and bottom panel of Fig.~\ref{mds_other}) the gas tends to cluster more clearly around $(90^\circ,0^\circ)$, which is the initial direction of the cloud orbit. This is because the dynamical interaction with the binary is not able to efficiently change the angular momentum direction of the surrounding gas on such short time-scales. In conclusion, the mini-discs that might arise from these perpendicular accretion events will be completely misaligned respect to the binary orbit, but they are likely to be roughly aligned with each other. 

Due to the large misalignment between the possible mini-discs and the binary, we expect other periodic effects to appear, like the Kozai--Lidov oscillations \citep{Kozai1962,Lidov1962} for example, where a test particle around one component of the binary periodically exchanges its inclination for eccentricity. Using hydrodynamical simulations, \citet{Martin2014} showed that this effect is also present in fluid discs. The time-scale for these oscillations is expected to be several times the binary orbital period -- \citeauthor{Martin2014} estimate $t_{\rm KL}\approx 17(0.35a/R_{\rm out})^{3/2} P_b$ for an equally-massive, circular binary and a mini-disc with surface density given by $\Sigma\propto r^{-1.5}$, which implies periods over 20 binary periods for mini-discs inside the Hill radius.
However, the changes on the disc inclination and eccentricity are large, which could have implications on processes such as the shaping of jets, the feeding onto the SMBHs and star formation.

In summary, as hinted by our set of simulations, the misalignment appears to be a natural outcome from  infalling cloud events, even when their orbits are aligned with the binary. Through the particular dynamics due to the interaction with the non-Keplerian gravitational potential of the binary, this could have important implications on the observability of these systems (see Section \ref{sec_fin}).

\section{Circumbinary discs}
\label{sec_cbd}

As we showed in Section \ref{sec_discs}, a circumbinary disk promptly forms \footnote{within the 20 orbits or so that we are modelling} whenever the impact parameter of the infalling cloud is large enough (or when its orbit is retrograde to that of the binary). However, our simulations are too short to assess the physical properties of these discs as they evolve toward a (possibly) steady state. In order to investigate the longer term evolution of these discs, we take two representative cases (namely, cases A3.0 and PE3.0 in table \ref{table_discs}). We re-simulate them for $\approx30$ further BHB orbits, using the final snapshots as initial conditions, but keeping only a sub-set of particles, to save computational time. We selected only particles bound to the binary with a period smaller than 30 times the binary period, as the contribution of the excised particles is dynamically negligible.

\begin{figure}
\centering
 \begin{picture}(230,490)
  \put(0,220){\includegraphics[width=0.46\textwidth]{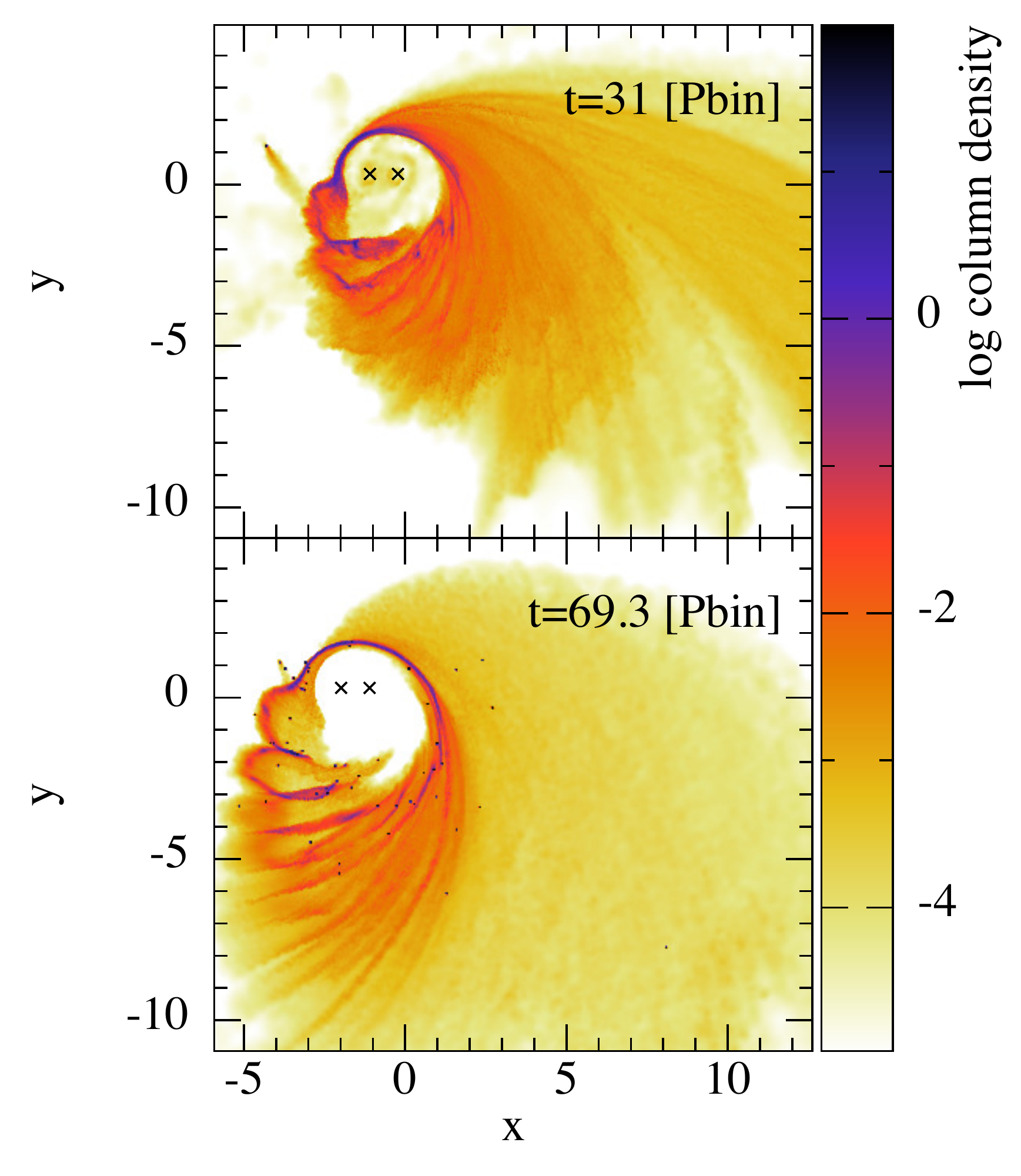}}
  \put(0,0){\includegraphics[width=0.46\textwidth]{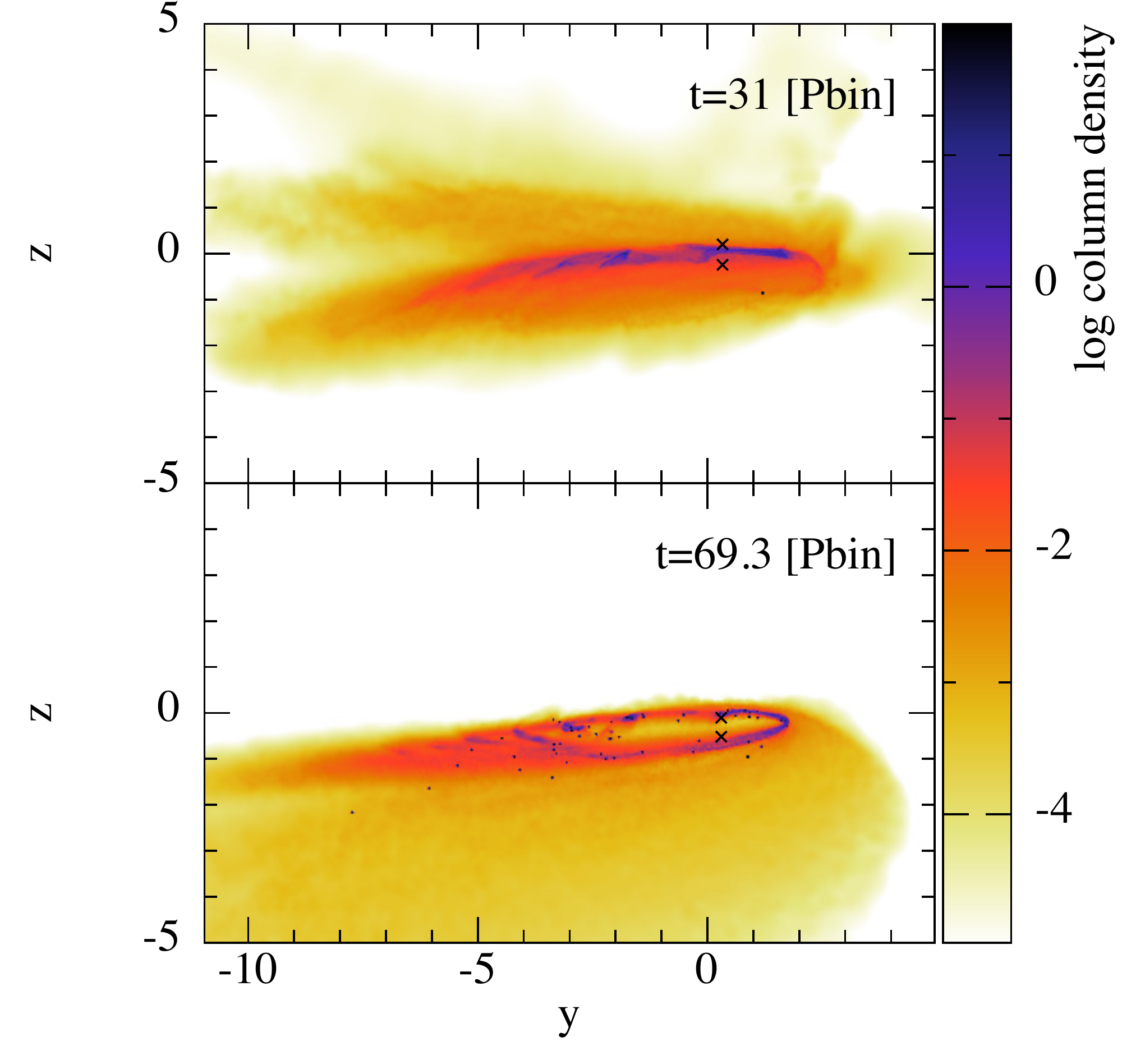}}
 \end{picture}
\caption{Column density maps of the PE3.0 re-simulation. The different times correspond to the beginning and the end of the re-simulation. The upper and lower panels show face-on and edge-on views of the disc, respectively.}
\label{resim_xz}
\end{figure}

\begin{figure}
\centering
\includegraphics[width=0.46\textwidth]{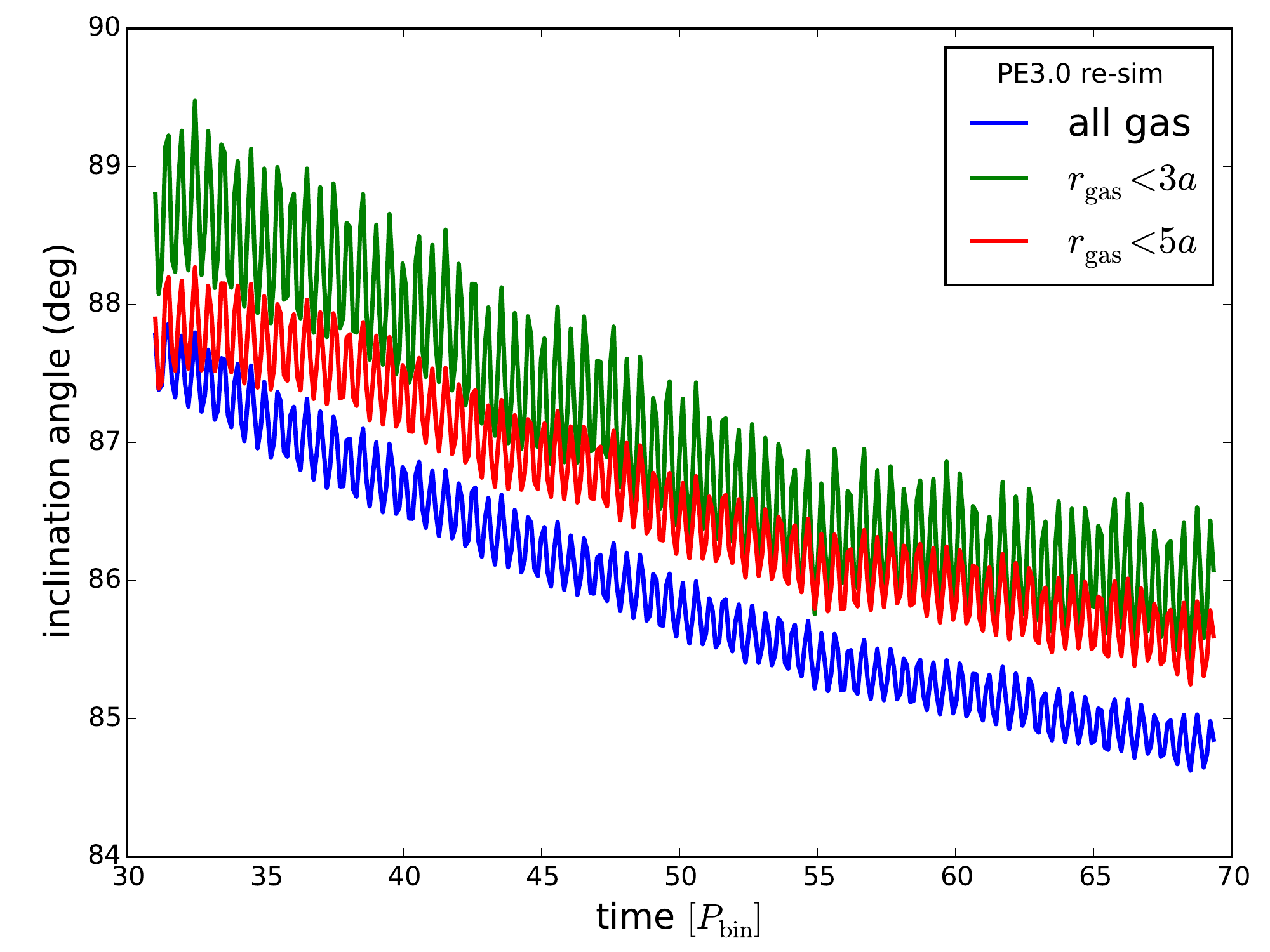}
\caption{Time evolution of the gas inclination on the re-simulation of the PE3.0 model. Because the circumbinary disc is not a well-defined structure, we sum the angular momentum of all the gas, and also  only that within two different radii, as indicated on the legend.}
\label{cbd1}
\end{figure}

\begin{figure}
\centering
\includegraphics[width=0.46\textwidth]{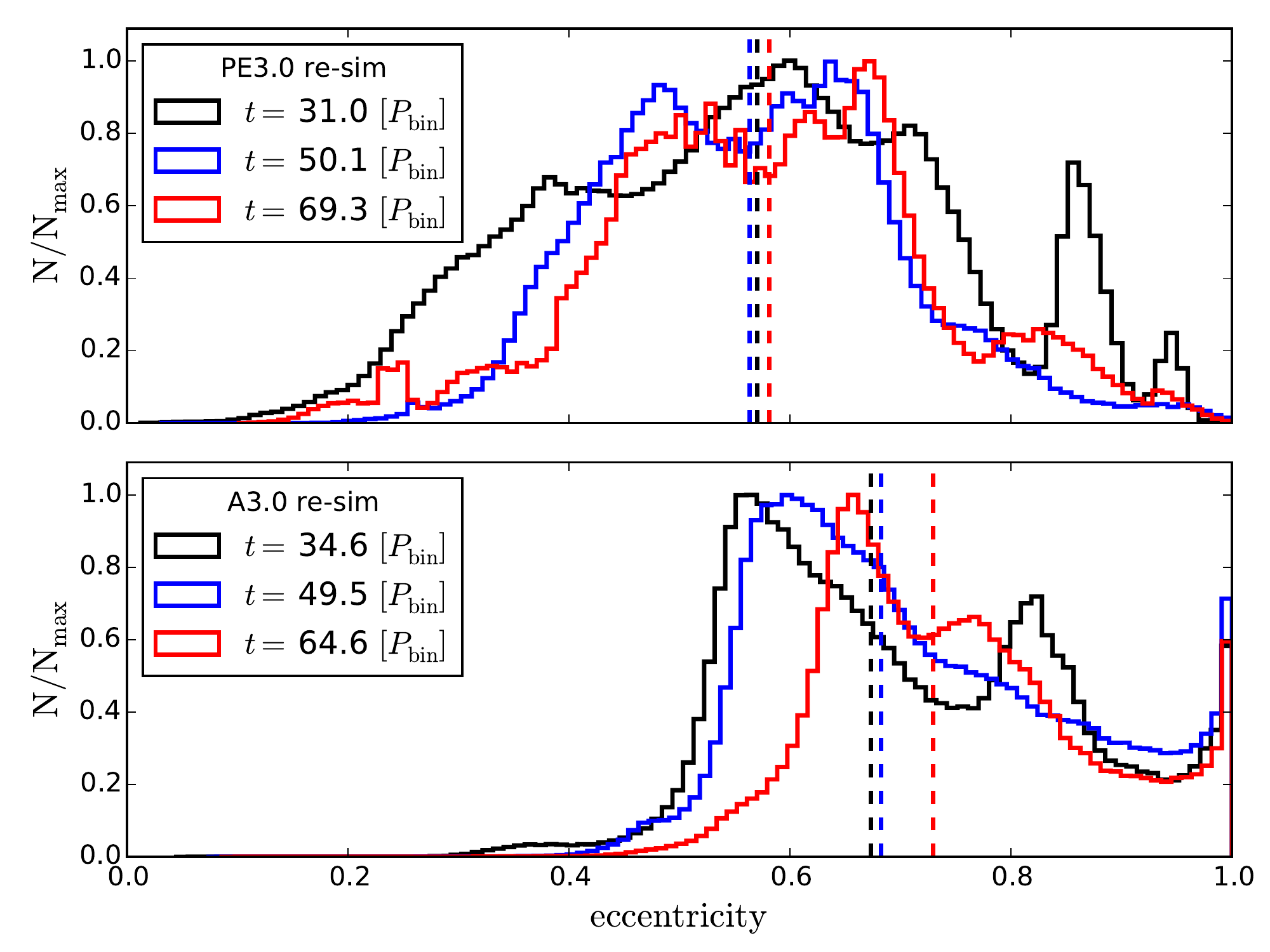}
\caption{Eccentricity distribution of the gas for PE3.0 (top) and A3.0 (bottom) re-simulations at different times. Each distribution is normalised by its maximum value. The vertical, dashed lines indicate the median of each distribution. For the perpendicular disc we observe that the distribution narrows, keeping roughly constant its median value, while for the aligned disc the distribution shifts towards higher values with time.}
\label{eccs_resim}
\end{figure}

The PE3.0 model is particularly interesting, because the circumbinary disc retains memory of the original orientation of the cloud and it is, therefore, perpendicular to the binary orbit. Column density maps at the beginning and the end of the re-simulation, from two edge-on views of the binary, are shown in Fig.~\ref{resim_xz}. In the lower panel, where the disc is seen roughly edge-on, we observe how it tends to slowly align to the binary orbit. In order to measure this evolution, we use the total angular momentum of the gas within a fixed radius as a proxy for the direction of the disc. The evolution of the inclination angle is shown in Fig.~\ref{cbd1} for different definitions of the disc extent, showing that this result is robust. We estimate that the alignment timescale is around 1000 orbits, should the evolution remain roughly constant. However, as explained for the mini-discs evolution, this timescale will depend critically on the viscosity treatment, and its study is beyond the scope of this paper.
We also notice that the inclination evolution shows the same oscillations that the mini-discs showed in the aligned case.  These oscillations have also half of the binary period, as expected since the dynamics is driven by similar processes as in the former case. This periodic perturbation has been overlooked on most studies of misaligned discs because it does not have a  secular effect on its evolution. However, our model suggests that it might have some interesting implications. For example, the density waves produced by the oscillatory perturbations enhance fragmentation on the gas, that we can see in the form of clumps in Fig.~\ref{cbd1}. Some of these clumps may form stars, a fraction of which will end up producing observable tidal disruption events (TDEs; see Section \ref{sec_fin}). Another possible signature of this oscillation might be imprinted in the shifting of spectral lines, specially coming from the inner regions where their amplitude is larger.
The longer simulation also allow to investigate trends in the circumbinary disc eccentricity. By inspecting the face-on views (upper panels of Fig.~\ref{resim_xz}) it seems that the material is more concentrated on eccentric orbits towards the end of the simulation. In order to measure this, we plot the eccentricity distribution at three selected times in the upper panel of Fig.~\ref{eccs_resim}. The distribution becomes narrower as the simulation advances, but actually keeps essentially the same median eccentricity of $\approx 0.6$.

\begin{figure}
\includegraphics[width=0.46\textwidth]{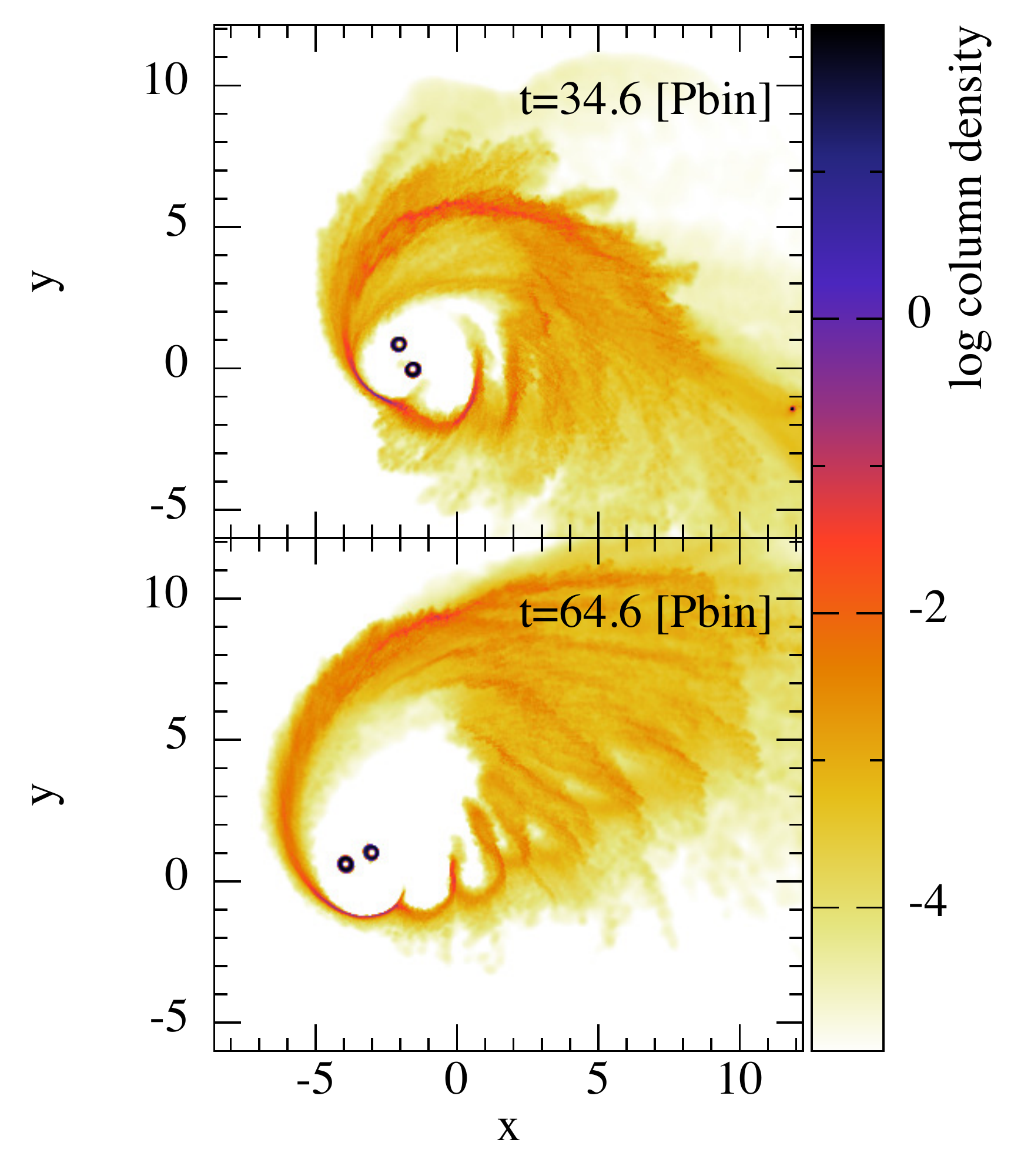}
\caption{Column density map at the beginning and the end of the A3.0 re-simulation. In this case we only show a face-on view of the disc and the binary, because they are almost completely aligned. Here we see that the disc increases its eccentricity.}
\label{resim_xy}
\end{figure}

For the re-simulated A3.0 model, column density maps for the beginning and the end are shown in Fig.~\ref{resim_xy}. As expected, the gas and the binary are essentially coplanar, with an initial inclination of only $\approx 4^\circ$, which decreases by $\approx 0.5^\circ$ during the re-simulation. The eccentricity evolution, shown in the lower panel of Fig.~\ref{eccs_resim}, is more interesting.  Here we observe a different behaviour respect to the model PE3.0; the orbits become more eccentric as time advances, which is reflected in the shifting of the distribution towards higher values and a clear increase of the median. 

\begin{figure}
\includegraphics[width=0.46\textwidth]{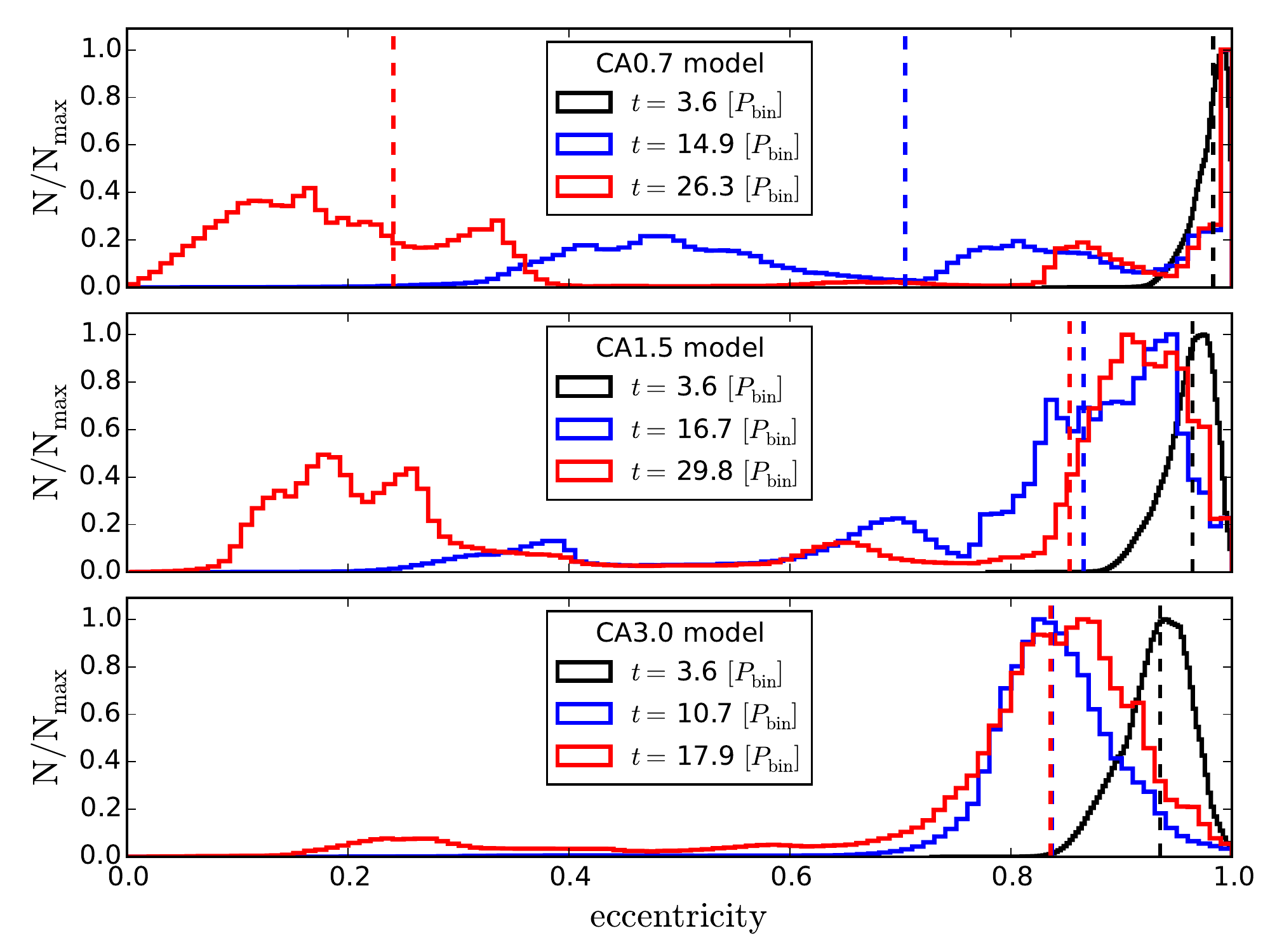}
\caption{Eccentricity distribution of the gas for the counter-aligned simulations at different times, increasing impact parameter from top to bottom. Each distribution is normalised by its maximum value. The vertical, dashed lines indicate the median of each distribution. In the three models we observe the material decreasing its eccentricity with time, reaching values close to zero. This process occurs faster for smaller impact parameters.}
\label{eccs}
\end{figure}

Finally,  a clear circumbinary ring also appears in our simulations of the counter-aligned cases, for all the investigated initial parameters (see upper middle row of Fig. \ref{BHB}). In this configuration, the formation of a clear circumbinary structure is extremely quick, and mass adds-up as the very eccentric tail joins it. We show the evolution of the gas eccentricity distributions in Fig.~\ref{eccs}. At the beginning the distribution is completely skewed towards high eccentricities due to the initial conditions, but after the interaction with the binary the gas eccentricity shifts towards lower values, and a striking bi-modality appears. This is clearly related to the formation of the ring, and it occurs on shorter time-scales for smaller impact parameters.

All our circumbinary discs appear to evolve differently according to their relative inclination to the binary. In the aligned model, the gas tends to increase its eccentricity; in the counter-aligned it tends to become circular, while in the perpendicular the eccentricity  retains its value.  
This is driven by the dynamical interaction with the binary:  a prograde encounter of a gas particle with one of the SMBHs will tend to increase its specific energy and angular momentum, increasing the eccentricity;  on the other hand, a retrograde encounter will work in the opposite direction, circularising the material.  For the perpendicular encounters, the binary is unable to change the orbital angular momentum of the gas, keeping its eccentricity roughly constant during the interaction.

\section{Physical scaling and observational consequences}
\label{sec_fin}

The efforts to confirm observationally the existence of SMBH binaries have increased in the last few years, motivated by their key role as probes of the hierarchical growth of galaxies \citep{Sesana2011}. Observing these objects is unfortunately very challenging, as their separations cannot be resolved by current capabilities on most galactic nuclei. Additionally, for the very few existent candidates, the observed signatures can also be explained by alternative scenarios that do not require a BHB (for a compilation of candidates and prospects in observational searches for BHBs, see \citealt{Dotti2012} and \citealt{Bogdanovic2014}). 

In the late stages of their evolution, when the binary separation is sufficiently low ($\lesssim 10^{-3}$ pc), gravitational waves will efficiently extract its angular momentum and energy, rapidly leading it to coalescence. The enormous amount of gravitational wave emission will be detectable by pulsar timing arrays \citep{Sesana2013} or by space-based missions like eLISA \citep{AmaroSeoane2013}.  However,  these detections will need to be complemented with observations in the electromagnetic spectrum in order to localise and characterise the sources.

\subsection{Physical scaling}
\label{sec:scaling}

In order to link our results with possible observational signatures we scale all physical quantities from code to physical units. 
To perform the scaling, we have to choose two parameters:  the mass of the binary and the critical density for the equation of state (eq.~\ref{rhoc}); fixing these two values will determine the rest of the units. 
\begin{table*}
\centering
\caption{Compilation of physical units of the initial values on our simulations. From left to right: binary mass, critical density for EoS, separation of the binary orbit, separation in terms of Schwarzschild radius, binary period, cloud mass, initial distance between the cloud and the binary, cloud radius, modulus of the cloud initial velocity, cloud velocity at the periastron for the smaller impact parameter, initial temperature of the cloud.}\label{units}
\begin{tabular}{l*{10}{c}c}
\hline\hline
 $M_{\rm bin}$ ($M_\odot$) & $\rho_{\rm cr}$ (g cm$^{-3}$) & $a$ (pc) & ${a/R_{\rm sch}}$ & $P_{\rm bin}$ (yrs) & ${M_{\rm cl}}$ ($M_\odot$) & ${R_{\rm cl}}$ (pc) & ${v_{\rm ini}}$ (km/s) & ${v_{\rm peri,0.7}}$ (km/s) & ${T_{\rm ini}}$ (K) \\
\hline
$10^6$ & $10^{-14}$ & $0.2$& $10^6$ & 8370 & $10^4$ & 0.5 & 40.8 & 342.9 & 100 \\
  $10^6$ & $10^{-12}$  & $0.04$& $2\times10^5$ & 837 & $10^4$ & 0.1 & 85.7 & 720 & 470 \\  
 $10^6$ & $10^{-10}$ & $0.009$& $5\times10^4$ & 84 & $10^4$ & 0.02 & 187.7 & 1577.3 & 2170 \\ 
 $10^7$ & $10^{-14}$ & $0.4$& $2\times10^5$ & 8370 & $10^5$ & 1 & 85.7 & 720 & 470 \\  
 $10^7$ & $10^{-12}$  & $0.09$& $5\times10^4$ & 837 & $10^5$ & 0.2 & 187.7 & 1577.3 &  2170\\  
 $10^7$ & $10^{-10}$ & $0.02$& $10^4$ & 84 & $10^5$ & 0.05 & 408 & 3429 &  $10^4$\\ 
 $10^8$ & $10^{-14}$ & $0.9$& $5\times10^4$ & 8370 & $10^6$ & 2.3 & 187.7 & 1577.3 & 2170 \\  
 $10^8$ & $10^{-12}$  & $0.2$& $10^4$ & 837 & $10^6$ & 0.5 & 408 & 3429 &  $10^4$\\  
 $10^8$ & $10^{-10}$ & $0.04$& $2\times10^3$ & 84 & $10^6$ & 0.1 & 877.2 & 7372.5 &  $4.6\times10^4$\\ 
\hline
\end{tabular}
\label{scaling}
\end{table*}

In Table \ref{units} we present physical scaling down to 84 years. Shorter periods would require too large densities compared to what is observed for molecular clouds. Moreover, for the more massive systems, a period of 84 years already implies a cloud temperature of $\sim 10^4$ K, which is already unrealistically high \citep[see e.g.][]{Meijerink2007} and cannot be pushed further. A period of roughly a century is too long to look for the variability associated to it in observed data.  Nevertheless, even if our model is unable to represent directly more compact systems, we still expect that the behaviour we find is qualitatively representative of those more rapidly varying systems, and below we discuss several possible observational signatures. In upcoming publications we will present models designed specifically for more compact systems.

\subsection{Observational signatures}

The feeding rates onto the binary and each SMBH show variability for all the configurations we model, always related with the orbital period. 
Recall that we measure the accretion rates at the sink radius $R_{\rm sink}$, which is large compared to the Schwarzschild radius.  Still, what AGN observations reveal is the luminosity of the accretion disc at different radii depending on the measured wavelength.
We expect the variable accretion rate we obtain at $R_{\rm sink}$ to represent actual variability of the accretion disc at large radii, which in some AGN shows up in optical or infrared light curves \citep[e.g.,][]{Lira2011}.

Variability due to binary feeding could thus appear in AGN light curves \citep[see][for the iconic case of OJ 287]{Sillanpaa1988}, and  be detected with future time-domain surveys like the Large Synoptic Survey Telescope \citep[LSST;][]{Ivezic2008}. 
Interestingly, periodicity has been recently claimed for a couple of systems proposed as binary candidates \citep{Graham2015,Liu2015}.

The presence of the mini-discs in our simulations, with their misalignment and evolution, allows us to explore novel and promising observational features.
\citet{Graham2015} showed that the blazar PG 1302-102  has a roughly sinusoidal light curve with a period of $\approx5$ years, and mentioned as a possible explanation the precession of a jet. 
If we assume that the orientation of a possible jet is given by the mini-disc\footnote{The jet orientation could also be given by the spin of the SMBH, which is not necessarily related with the mini-disc orientation, specially at these scales.}, we expect to observe variability related to the mini-disc wobbling found in our models and predicted earlier by \cite{Bate2000}, specially if the jet is close to the line of sight.  \citet{Graham2015} fitted the observed light curve with a wobbling amplitude of around $0.5^\circ$ which is actually very close to what we find in Fig.~\ref{precess}. Therefore, we suggest that the observed variability of this source could be due to mini-disc wobbling as shown in our models. Notice that this interpretation would imply a rest-frame orbital period for the binary of eight years, one order of magnitude longer than under the assumption that the variability is due to lumps in the circumbinary disc \citep{DOrazio2015}.  

In the cases where the viscosity of the individual discs is low, the differential precession induced by the companion SMBH is not communicated efficiently throughout the discs, which then could break. The disruption of the disc into rings that precess independently could also be a source of variability, as the dissipation between the gaseous rings is enhanced and this promotes stronger accretion onto the central SMBH \citep{Dogan2015}.  There is recent observational evidence for this process in a proto-planetary disc \citep{Casassus2015}.

Additional observable features could be due to warps in the misaligned mini-discs. We do not find warps in our simulations due to the relatively low resolution we can afford.  However, they are expected due to the gravitational pull of the companion \citep{Moeckel2006}. 
The presence of a warp can cause part of the disc to block other parts of the disc or a central source.  Additionally, it can change the viewing angle which is important for synchrotron emission.  Both effects occur periodically, on the dynamical time-scale of the disc.
The effects of warps have been observed in several astrophysical objects such as active galactic nuclei, circumstellar discs and stellar mass black holes through variability in (i) the photometry \citep[e.g.][]{Herrnstein2005,Manset2009,Bouvier2013}; (ii) the spectrum \citep[e.g.][]{Reynolds2009,Looper2010}; and (iii) polarisation \citep[e.g.][]{Manset2009,Roland2009,Cheng2015}.

Circumbinary discs can also have variable emission. In particular, Figs. \ref{resim_xz} and \ref{resim_xy} show a series of discrete shock fronts propagating from the binary into the tail of the eccentric disc. Such regular feature is not seen in comparable simulations featuring a single SMBH \citep{BR08}. Material approaching the periastron is accelerated and flung away into the tail when it is in phase with one of the two SMBHs. This creates ejected waves with a periodicity of half the binary orbital period that compress and shock the surrounding material, possibly leading to periodic enhancements in luminosity and discrete episodes of star formation. As the gas is located farther from the black holes ($r\gtrsim 2a$, for a prograde disc), we expect the period of such variability to be typically longer than the binary period \citep[e.g.][]{Farris2015}. Then, a source with a photometric period shorter than the spectroscopic period could be interpreted as a binary surrounded by a disc -- the binary varies on one, or half, an orbital period due to accretion, while the circumbinary varies on its own dynamical time.
Moreover, the ratio between both variability periods could be a tool to study the geometry and extension of the gas around the binary.

We have shown that increasing the impact parameter results in more clumping for the gas. The presence of clumps could have indirect implications on the observability of a binary through star formation and posterior tidal disruption events (TDEs). \citet{Pau2013}  showed that  fragmentation in a circumbinary disc results on {\em in-situ} star formation and an increase of the rate of TDEs respect to what is expected for typical galaxies. Later on, \citet{Brem2014} demonstrated that the presence of the binary instead of a single SMBH will produce a distinctive signature on the reverberation mapping after such an event. Based on this, we might expect to detect these circumbinary structures through TDEs. Although the rate will depend on the efficiency of star formation on the gas tail and the disc itself, which we do not model here. In any case, LSST will detect thousands of TDEs, greatly increasing the chances of detecting this kind of events even if they are relatively rare.

Finally, star formation in a circumbinary disc could also refill the loss-cone and affect the evolution of the binary orbit by exchange of energy and angular momentum with the stars via 3-body interactions \citep{Pau2013}. Moreover, eccentric discs of stars, which is the most likely output of the near-radial infall of clouds that we are modelling (Figs. \ref{eccs_resim} and \ref{eccs}), will be subject to instabilities  \citep[see e.g.][]{Madigan2009} that can increase even further the amount of stars with orbits that will interact directly with the binary, enhancing the probability of TDEs and/or evolving the binary orbit.

\section{Summary}
\label{sec:summary}

We presented numerical, SPH models of the evolution of turbulent clouds in near-radial infall onto equal-mass, circular supermassive black hole binaries. In order to explore different orbital configurations for the cloud we performed a total of 12 simulations, changing both the impact parameter and the relative inclination between orbits.

We studied the formation of discs around the binary and each SMBH depending on those orbital configurations. Our main findings can be summarised as follows:
\begin{enumerate}
\item We only observe the formation of stable and prominent (up to the Hill radius) mini-discs for the aligned models, independent of the cloud impact parameter.  For the other orientations, where the gas trapped by the individual SMBHs has little angular momentum, we do not have the spatial resolution to observe the possible formation of smaller mini-discs.
\item The misalignment of the mini-discs around each SMBH seems to be a natural outcome of the infall of extended gas clouds.  We have confirmed the analytical result that misaligned mini-discs will precess and wobble due to the presence of a companion SMBH.
\item When the impact parameter is large enough (pericentre distance $\approx 3R_{\rm bin}$) we always observe the formation of a circumbinary disc, independent of the orbit orientation. In the counter-aligned cases there is formation of a circumbinary disc (or ring) for all impact parameters.
\item The circumbinary discs tend to follow the initial orientation of the cloud. We have confirmed that a misaligned disc will evolve towards alignment.
\item The circumbinary discs are initially very eccentric ($e\gtrsim 0.6$), and their early evolution will depend on the relative orbital inclination. In the aligned model, the gas tends to increase its eccentricity, while the opposite is true in the counter-aligned model. In the perpendicular cases the eccentricity distribution becomes narrower without changing its median value.
\item The feeding rates onto the binary and each SMBH show variability for all our models, always related to the orbital period.
\end{enumerate}

Although our simulations do not allow a direct scaling to BHBs with orbital periods of few years (i.e., accessible to future time-domain surveys), we argue that the observed qualitative phenomenology, which is mostly driven by the gravitational torques exerted by the BHB onto the gas, can be safely extrapolated to shorter periods. Such phenomenology has distinctive electromagnetic signatures, which are very relevant for the future identification and characterisation of BHBs.

\section*{Acknowledgments}
We thank the anonymous referee for comments that helped improving the clarity of this paper.
We also thank Johanna Coronado and Alex Dunhill for their comments on the manuscript; Carlos Cartes for his assistance with the turbulent initial conditions for the cloud;  Daniel Price, Ralph Klessen and Paul Clarke for their helpful comments on some numerical issues; Xian Chen for pointing out the different variability time-scales for the BHB accretion and the circumbinary spectrum;  William Lucas for his assistance with the initial conditions for the cloud orbit; and Patricia Ar\'evalo for useful discussions on AGN variability.

F.G.G and J.C. acknowledge the kind hospitality of the AEI and the AIP.

Column density maps were created with {\sc splash} by \cite{Splash}. The simulations were performed on the {\it datura} cluster at the AEI, and also on the {\it geryon} computers at the Center for Astro-Engineering UC (BASAL PFB-06, QUIMAL 130008, Fondequip AIC-57).

We acknowledge support from CONICYT-Chile through FONDECYT (1141175), Basal (PFB0609), Anillo (ACT1101), Redes (120021), and Exchange (PCCI130064) grants; from the DAAD (57055277); and from the European Commission's Framework Programme 7, through the Marie Curie International Research Staff
Exchange Scheme LACEGAL  (PIRSES-GA-2010-269264).
F.G.G. is supported by CONICYT PCHA/Doctorado Nacional scholarship. F.S. is supported by the DFG project FOR1254. A.S. is supported by a University Research Fellowship of the Royal Society.

\bibliographystyle{mnras}
\bibliography{refs}

\bsp

\label{lastpage}

\end{document}